\documentclass[twocol]{ametsocV5}

\title{A Bayesian approach to regional decadal predictability: Sparse parameter estimation in high-dimensional linear inverse models of high-latitude sea surface temperature variability}

\authors{Dallas Foster\correspondingauthor{Dallas Foster, fostdall@oregonstate.edu}}

\affiliation{Oregon State University, Corvallis, Oregon 97331, USA \\
	Computational Physics and Methods (CCS-2), Los Alamos National Laboratory, Los Alamos, NM 87545, USA.}

\extraauthor{Darin Comeau}
\extraaffil{Computational Physics and Methods (CCS-2), Los Alamos National Laboratory, Los Alamos, NM 87545, USA.}

\extraauthor{Nathan M. Urban}
\extraaffil{Computational Physics and Methods (CCS-2), Los Alamos National Laboratory, Los Alamos, NM 87545, USA.}

\usepackage{multirow}
\usepackage{amsmath,amsfonts,amssymb,bm}
\usepackage{mathptmx}
\usepackage{newtxtext}
\usepackage{newtxmath}
\usepackage{anyfontsize}
\usepackage{hyperref}

\newcommand\blfootnote[1]{%
  \begingroup
  \renewcommand\thefootnote{}\footnote{#1}%
  \addtocounter{footnote}{-1}%
  \endgroup
}

\newcommand{\x}{\pmb{x}}
\newcommand{\y}{\pmb{y}}
\newcommand{\R}{\mathbb{R}}
\newcommand{\B}{\pmb{B}}
\newcommand{\Q}{\pmb{Q}}
\newcommand{\G}{\pmb{G}}
\newcommand{\Sig}{\pmb{\Sigma}}
\DeclareMathOperator*{\argmax}{arg\,max}

\abstract{Stochastic reduced models are an important tool in climate systems whose many spatial and temporal scales cannot be fully discretized or underlying physics may not be fully accounted for.  One form of reduced model, the linear inverse model (LIM), has been widely used for regional climate predictability studies - typically focusing more on tropical or mid-latitude studies. However, most LIM fitting techniques rely on point estimation techniques deriving from fluctuation-dissipation theory. In this methodological study we explore the use of Bayesian inference techniques for LIM parameter estimation of sea surface temperature (SST), to quantify the skillful decadal predictability of Bayesian LIM models at high latitudes.  We show that Bayesian methods, when compared to traditional point estimation methods for LIM-type models, provide better calibrated probabilistic skill, while simultaneously providing better point estimates due to the regularization effect of the prior distribution in high-dimensional problems. We compare the effect of several priors, as well as maximum likelihood estimates, on (1) estimating parameter values on a perfect model experiment and (2) producing calibrated 1-year SST anomaly forecast distributions using a pre-industrial control run of the Community Earth System Model (CESM). Finally, we employ a host of probabilistic skill metrics to determine the extent to which a LIM can forecast SST anomalies at high latitudes. We find that the choice of prior distribution has an appreciable impact on estimation outcomes, and priors that emphasize physically relevant properties enhance the model's ability to capture variability of SST anomalies. }

\begin{document}

\maketitle

\section{Introduction} \label{sec:INTRO}
\blfootnote{\textbf{This work has been accepted to the Journal of Climate. The AMS does not guarantee that the copy provided here is an accurate copy of the final published work.}} Inter-annual variability in sea surface temperatures (SSTs) is an important indicator of the variability of the global climate system and has long been studied as relating to other atmospheric variables \citep{Davis1976, McKinnon2018, Revelard2018}. Forecasting variations in SSTs themselves has been an area of open research: there is interest not only in the time scale of predictability \citep{Davis1976, Branstator2010, Guemas2012, Stock2015}, but also on regional phenomena such as the El-Ni\~no-Southern Oscillation (ENSO) and the Pacific Decadal Oscillation (PDO) \citep{Thomas2018, DiNezio2017, Penland1996, Penland1993, Hawkins2009, Wittenberg2014} which can have global impacts through atmospheric teleconnections. Stochastic reduced models, like the linear inverse model (LIM), have been widely used for regional climate predictability studies. A LIM is a linear stochastic differential equation model for the evolution of a climate field, such as SSTs. Often, and is done in this paper, the empirical orthogonal functions (EOFs) of said field are modeled (see \citet{Penland1989} for a discussion of the use of EOFs in LIMs). The most prominant application of LIMs are to sub-annual prediction of SSTs for ENSO forecasting \citep{Penland1993, Penland1995}, but there are also applications to decadal timescales and mid-latitude regions \citep{Alexander2008, Newman2007, Newman2013a, Dias2018, DelSole2013}, extensions to systems with nonlinear and memory effects \citep{Kravtsov2009a, Kondrashov2015a, Chenb}, as well as a variety of other contexts \citep{Hawkins2009, Penland2001, Wu2018, Martinez2018, Alexander2008, Barnston1999}.

The high-latitudes have long been the subject of numerous climate predictability studies, with Arctic sea ice in particular garnering considerable attention. While practical needs drive predictions of Arctic sea ice on subseasonal to seasonal timescales, many studies have also pushed the forecast horizons to interannual to decadal timescales (\cite{guemas2016review} and references therein). In general, interest in decadal climate predictions has been increasing over the last several years \citep{meehl2014decadal}, including focus on SSTs \citep{hawkins2011evaluating} and high latitudes, particularly in the Southern Ocean \citep{latif2017southern, zhang2017diagnosis}, where interannual to decadal scale variability plays a strong role in oceanic forcing on the Antarctic ice sheet, particularly to the vulnerable West Antarctic Ice Sheet \citep{jenkins2016decadal}.

The main focus in this paper is the implementation, calibration and evaluation of the LIM framework as applied to the forecasting of high latitude SST anomalies using modern Bayesian statistical strategies and probabilistic scoring. We make use of the 1800 year fully-coupled pre-industrial control run of the Community Earth System Model version 1 (CESM1) Large Ensemble project \citep{EKaya}, and was chosen to ensure that there was enough data to perform validation of the methodology and to avoid complicating factors of non-stationarity, data sparsity or bias. We adopt a Bayesian perspective because of the regularization effect that the use of prior distribution have in high-dimensional estimation problems \citep{Hastie2017,  Vogel2002} and the ability to produce better probabilistic calibration when compared to point estimation \citep{Samaniego2010}.

The parameters of a LIM, i.e. the drift and diffusion matrices, must be inferred from data. The inference of the drift term is traditionally, e.g. in \citep{Penland1989}, achieved by using fluctuation-dissipation theory (FDT) relating the matrices and the data covariances. Estimation of the drift and incremental noise covariance matrices in this way are also equivalent to the Maximum Likelihood (ML) estimates as also shown in \citep{Penland1989} and presented in section \ref{sec:MLE}. To simplify the estimation problem, the dynamics of the system are assumed linear and Markovian so that the eigenfunctions -- or so called principal oscillation patterns (POPs) -- of the propagating matrices can be analyzed for spatial patterns \citep{vonStorch1988, Hasselman1988}. This technique produces point estimates for the parameters which allows for scientists to infer the physical characteristics of the EOFs from the system parameters (i.e. the spectrum of the drift matrix) and to forecast climate statistics using a reduced model. ML and Frequentist approachess can be used to obtain sample statistics like confidence intervals, and skewness and kurtosis estimates, as in \citet{Martinez2018}, which provides insight to parameter uncertainty but we argue that the Bayesian perspective provides a complete framework to specify prior information and propagate parameter uncertainty through the forecasting model. 

To facilitate the main focus of this paper, the secondary objective is to consider the LIM problem from a Bayesian perspective and explore implementation strategies. Bayesian parameter estimation diverts from prevailing frequentist statistical estimation by defining the parameters as random variables with probability distributions and the subsequent role of diverse prior probabilities given to the estimands. For a brief review of Bayesian techniques in parameter estimation see \citet{Gelman2013}. Notably, filtering approaches, as, for example, in \citet{Hansen2007a} and \citet{Zhao2019}, offer convenient approaches to state and parameter estimation at the cost of the assumption of Gaussian or particle filter distributions. The filtering approach can be beneficial for time-dependent parameters, but may also have difficulty with filter divergence or with exploring the entire posterior probability space \citep{Sarkka2013}. Because the LIM parameters are not time dependent, we have chosen in this paper a MCMC variant approach to estimate the posterior distribution of the parameters.

In our application there are two primary motivations for a Bayesian approach to parameter estimation. First, we argue that the likelihood distribution induced by a LIM is relatively weak, so that the posterior distribution is highly sensitive to a prior distribution and this sensitivity leads to improvements in parameter recovery and forecast skill. In other words, the problem is high-dimensional and the relative amount of information per parameter is relatively low. The introduction of prior densities in the Bayesian framework is a way to add additional information to, or regularize, the parameter distributions. It is known that, in parameter estimation, the Bayesian estimates will outperform the frequentist estimate as long as the Bayesian prior does not \textbf{both} have large error in mean and relative large weight \citep{Samaniego2010}. Second, we argue that propagating parameter uncertainties through the model can produce better calibrated forecast distributions. In the traditional ML approach, only the point estimates are typically used to produce forecast distributions, although cross-validation methods provide another way to regularize parameter estimates and produce an ensemble of forecasts. The Bayesian approach provides a framework and means to characterize the distribution of parameter uncertainties and their effect on the model and forecast distribution. By fully characterizing this uncertainty, we believe that the Bayesian methods detailed in this paper are better suited to reproduce observation distributions when compared to a traditional point-estimate approach. 

The third thrust of this paper is a focus on the performance of the Bayesian perspective (in relation to MLE/FDT) as a function of the specific prior implemented. Using nontrivial priors in time series analysis has a long history in economics, where Bayesian methods are used to estimate coefficients (often) of Vector Auto-Regressive (VAR) models \citep{Litterman1986a, Giannone2015, Partridge1998, Doan1983}. There have been various attempts to introduce Bayesian learning of parameters in stochastic differential equations, but usually require some assumptions or approximations of the underlying likelihood distribution \citep{Eraker2001, Beskos2006, Karimi2016, Batz2017, Tian2014, Singer, Mller2011}. Introduction of nontrivial prior beliefs about the parameters can complicate the process of describing the necessary parameters since the densities resulting from stochastic differential equations do not generally have closed form expressions. Consideration of complex prior distributions in parametric models is underdeveloped, but the effect of prior densities in non-parametric modeling has been studied in \citet{VanZanten2013}.  

Statistical estimation of stochastic models has a long history starting with ML and least square estimates as a simplification, see \citet{LeBreton1977} and \citet{DelSole2010}. In application to climate statistics, reviews can be found in \citet{DelSole1999} and \citet{DelSole2003} and comparisons in \citet{Mason2002, Nummelin2018}. From \citet{Penland1989}, the use of FDT approximations for the LIM to derive closed-form parameter estimation formulas makes this approach the method of choice in geoscience applications \citep{Penland1989, Penland1995, Nummelin2018, Barnston1999, Colman2003a}, particularly over various autoregressive models \citep{DelSole2003, Penland1993}. \citet{Barnston2012a} compares LIMs and other physical and statistical based models in IRI ENSO prediction plume. In this paper, we use a statistical framework to show that the same estimates can be achieved from maximizing a natural parameter likelihood function and that the introduction of prior distributions in a Bayesian sense is a natural extension.

The regression interpretation of LIMs gives us reasoning to believe why regularization is necessary. The Gauss-Markov Theorem ensures that the MLE for LIMs have the smallest variance of all unbiased estimates, it is known in general that these estimates can be improved by various methods of regularization like Ridge Regression, LASSO and cross-validation that reduce estimator variance \citep{Hastie2017}. The need for regularization may be the result of the presence of confounding variables or model error and the lack of data relative to the number of parameters. In fact, the use of a Bayesian prior can be seen as a type of regularization that aims to address these limitations. 

In what follows we briefly discuss the theoretical underpinnings and results of LIMs, MLE for these models, and Bayesian Methods for Parameter Estimation. The aim of this paper is then to provide insight into the use of a Bayesian framework with nontrivial prior distributions and how they can be used to generate well-calibrated SST anomaly forecasts. To this point, in the results section, we give two experiments to compare the use of ML and Bayesian estimation methods. First, in a perfect model methodology, we compare convergence of parameter estimations. Second, we apply these methods to a LIM of SST anomalies and analyze the forecast performance of each using a variety of traditional and probabilistic skill metrics. We find that the use of Bayesian methods with informative priors is significantly more skillful in forecasts and can better estimate the underlying probability distribution of the data.

\section{Linear Inverse Modeling} \label{sec:LIM}
Consider the linear stochastic differential equation,
    \begin{equation}\label{eq:LSDE}
        d\x_t = \B\x_t dt + d\pmb{W}_t,
    \end{equation}
where $\x\in \R^{m}$, $\B\in R^{m\times m}$ and $\pmb{W}$ is an $m$-dimension Wiener process with zero mean and covariance given by $\Q \in \R^{m\times m}$. We denote the covariance of the process $\x$: $\langle\x \x^T\rangle = \pmb{\Lambda}$. The values of $\B$ or $\Q$ are not known a priori and the fitting of a LIM requires the estimation of these $m^2 + \frac{1}{2}m(m+1)$ total parameters, since $\Q$ is symmetric. 

The solution to (\ref{eq:LSDE}) at a time $t+\tau$ is given by
    \begin{equation}\label{eq:LSDEsolution}
        \x_{t+\tau} = \pmb{G}(\tau)\x_t + \boldsymbol{\xi}_t, \qquad \boldsymbol{\xi}_t \sim N(0, \Sig(\tau))
    \end{equation}
where 
    \begin{equation}\label{eq:LIMG}
    \G(\tau) = e^{\B \tau}
    \end{equation}
    is refered to as the propagating matrix of lead time $\tau$, but in the literature it is also called the Green's function for this differential equation, the incremental noise process satifies the relation
    \begin{equation}\label{eq:LIMSIG}
    \Sig(\tau) = \pmb{\Lambda} - \G(\tau) \pmb{\Lambda} \G(\tau),
    \end{equation} 
    and the fluctuation-dissipation relation gives
    \begin{equation}\label{eq:FDR}
        \B^T \pmb{\Lambda} + \pmb{\Lambda}^T\B = -\Q
    \end{equation} (see \citet{Penland1989} for detailed calculations). Both $\Q$ and $\Sig$ must be symmetric positive definite, while $\B$ must have eigenvalues with negative real part in order for the solution (\ref{eq:LSDEsolution}) to be stable. See \citet{Penland1995} for a lengthier discussion on tests for the applicability of the assumptions of LIMs and resulting approximations.

\section{Maximum Likelihood Estimation} \label{sec:MLE}
Now consider observations of the random process $\y_n = \x(t_n)$ for $n = 1, \dots, T$, abbreviated as $\y_{1:T}$ when considering a range of observations. We assume that the observation process is exact, but the introduction of observational noise can be accommodated with only minor adjustments to the methods detailed below.  The conditional probability density function, typically called the conditional marginal likelihood, for these observations conditioned on matrices $\G$ and $\Sig$ defined above is Gaussian and given by
    \begin{equation}\label{eq:CondLikelihood}
     p\left(\y_{1:T} \biggl\vert \G, \Sig \right) = 
    \prod_{n=1}^{T-\tau} \mathcal{N}\left(\y_{n+\tau} - G(\tau) \y_{n}, \Sig(\tau) \right) 
    \end{equation}
Taking the derivative of the marginal likelihood with respect to $\G(\tau)$ and $\Sig(\tau)$, we can derive formulas for the so-called maximum likelihood (ML) estimates
    \begin{align} \label{eq:MLEG}
        \hat{\pmb{G}}(\tau) &= \left(\sum\limits_{n=1}^{T-\tau} \y_{n+\tau} \y_{n}^T\right)\left(\sum\limits_{n=1}^{T-\tau} \y_{n}\y_{n}^T\right)^{-1}, \\ 
        \hat{\pmb{\Sig}}(\tau) &= \frac{1}{T-\tau-1} \sum_{n=1}^{T-\tau} \left(\y_{n+\tau} - e^{\B\tau} \y_{n}\right)\left(\y_{n+\tau} - e^{\B\tau} \y_{n}\right)^T \nonumber\\
        &= \langle \y_{\tau:T}\; \y_{\tau:T}^T \rangle - \hat{\G}(\tau) \langle \y_{1:T-\tau} \;\y_{1:T-\tau}^T\rangle \hat{\G}(\tau)^T \label{eq:MLESIG}
    \end{align}
From these matrices, one calculates the matrices $\hat{\B}$ using (\ref{eq:LIMG}) and $\hat{\Q}$ using (\ref{eq:FDR}) by approximating $\pmb{\Lambda}$ with ensemble averages. While $\hat{\Q}$ is formally obtained from the fluctuation dissipation relation, we will also refer to this as the ML estimate of $\Q$ since it is obtained from using the ML estimate of $\B$. Note that (\ref{eq:MLESIG}) is approximately equal to (\ref{eq:LIMSIG}) under the assumption of stationarity. 

\begin{figure*}[t]
    \noindent \includegraphics[width=39pc]{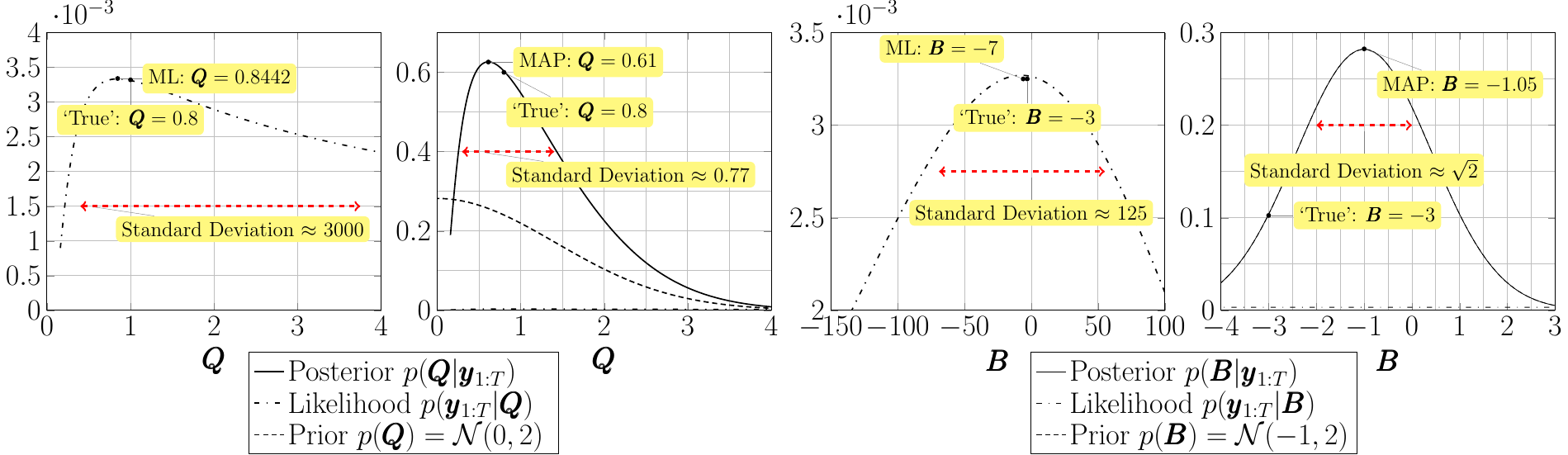}
    \caption{A stochastic process with known $\B$ and $\Q$ was generated by discretizing (\protect\ref{eq:LSDE}) using Euler-Maruyama with time step $dt$. 1000 observations from that series were sampled with $\tau = 100 dt$ to form $y_{1:T}$. From left to right: (1) trace plots of the marginal likelihood $p(\y_{1:T}|\Q)$, (2) marginal likelihood $p(\y_{1:T}|\Q)$ along with prior normal $p(\Q)$ and resulting posterior distribution $p(\Q| \y_{1:T})$, (3) marginal likelihood $p(y_{1:T}|\B)$, and finally (4) the marginal likelihood $p(\y_{1:T}|\B)$ along with prior normal $p(\B)$ and resulting posterior distribution $p(\B| \y_{1:T})$ are shown. Maximum Likelihood and Maximum A Posteriori are shown for comparison between Frequentist and Bayesian point-estimates while standard deviations of the distributions are shown to compare uncertainties in the parameters from the Bayesian perspective} \label{fig:MLE_MAP_Example}
\end{figure*}

As a numerical example, consider the 1 dimensional case. In \textbf{Fig.~(\ref{fig:MLE_MAP_Example})} a stochastic process $\x_t$ was generated using chosen values $\B = -3$ and $\Q = 0.8$ and sampled to generate 1000 observations $\y_k$. We then compute $p(\y_{1:T}|\G, \Sig)$, where we simply use $\tau=1$. For the scalar case, trace plots of $p(\y_{1:T}|\B, \Q)$ can be obtained from $p(y_{1:T}|\G, \Sig)$ since $\Sig$ and $\Q$ have the simple relationship $\Sig = (e^{2\B\tau}-1)\Q/2\B$ from solving (\ref{eq:LIMSIG}) and (\ref{eq:FDR}).  The traceplots for $\Q$ and $\B$ are displayed in the first and third image respectively, adjoined by their Bayesian estimates as described further in section \ref{sec:BAYES}\ref{sec:Priors}. The location of the maximum of these trace plots give ML estimate $\hat{\B}$, and $\hat{\Q}$, whose experimental relative error was found to be approximately  146\%, and 6\% respectively, while the relative error for $\hat{\G}$ was approximately $0.5\%$. Since the transformation $\B = \log{(\G)}/\tau$ increases the variance of the marginal likelihood and amplifies the relative error, it is conceivable from this experiment alone that reasonable forecasts results can be achieved with $\hat{G}$ while the numerical value of $\hat{B}$ differ significantly from the 'true' value $\B$. Additionally, observe that the marginal likelihood places significant probability on values of $\B$ which are positive, something which we know leads to unstable dynamics and is physically impossible. Regularization of the marginal likelihood can help constrain estimates from undesirable regions; the Bayesian approach can be seen as a particular justification of this regularization.

\section{Bayesian Methods for Parameter Estimation} \label{sec:BAYES}
Determining the most likely values and distributions for the LIM parameters is inherently a high-dimensional problem with hundreds, if not thousands, of parameters needing estimation. Because the ratio of the number of parameters to data points may be close to or larger than unity, regularization is needed to constrain possible parameter values and prevent over-fitting. In the statistical sense, regularization provides a way to balance bias and variance in the estimates. A Bayesian probabilistic framework provides a mechanism to formally regularize the maximum likelihood distribution by incorporating prior beliefs about the uncertainty of the parameters that is equivalent to Tikhonov regularization \citep{Vogel2002}, also known as ridge regression.

Bayesian philosophy assumes that the parameters $\theta \in \R^{d}$ are random variables with joint prior probability distribution $p(\theta)$. This prior distribution represents any a priori knowledge we have about the nature and structure of the parameters. A state space model with model states $x(t_k)$, observations $y_k$, and unknown parameters may be written as 
    \begin{equation} \label{eq:StateSpace}
    \begin{aligned}
        \theta \sim & \; p(\theta), \\
        \x(t_0) \sim & \; p(\x(t_0), \theta), \\
        \x(t_n) \sim & \; p(\x(t_n)| \x(t_{n-1}), \theta), \\
        \y_n \sim & \; p(\y_n | \x(t_n), \theta),\\
    \end{aligned}
    \end{equation}
where we label the initial uncertainty $p(\x(t_0), \theta)$, the transition probability density $p(\x(t_n)| \x(t_{n-1}), \theta)$, and observation uncertainty $p(\y_k|\x(t_k), \theta)$. To simplify the above relations, it is assumed that the initial condition and observation uncertainty are independent of the parameters $\theta$. The full posterior distribution, with these assumptions, can be written using Bayes' rule:
    \begin{equation}\label{eq:FullPosterior}{\scriptstyle
        p(\x(t_{0:T}), \theta | \y_{1:T}) = \frac{p(\x(t_0)) p(\theta)}{p(\y_{1:T})} \displaystyle\prod_{n=1}^{T} \scriptstyle p\left(\y_n | \x(t_n)\right)\; p\left(\x(t_n)|\x(t_{n-1}), \theta\right)}
    \end{equation}
Note that (\ref{eq:FullPosterior}) gives the joint posterior distribution for both the state $\x_{0:T}$ and the parameters $\theta$ and estimating it solves not only the parameter estimation problem but also the filtering/smoothing problem. The distribution of only the parameters requires marginalizing out the state variables,
    \begin{equation}\label{eq:Marginal}
        p(\theta | \y_{1:T}) = \int p(\x(t_{0:T}), \theta | \y_{1:T}) d\x(t_{0:T}).
    \end{equation}
Computation of this integral can be difficult, but possible using Monte Carlo methods; some possibilities are presented in the appendix. We make the standard approximation that the marginal distribution is given by
    \begin{equation}\label{eq:MarginalApprox}
        p(\theta | \y_{1:T}) \propto p(\y_{1:T} | \theta) p(\theta).
    \end{equation} 

Taking the maximum of (\ref{eq:MarginalApprox}) yields the Maximum A Posteriori (MAP) point estimate, $\hat{\theta}^{\textrm{MAP}}$,
    \begin{equation}\label{eq:MAP}
        \hat{\theta}^{\textrm{MAP}} = \arg \max \left[ p(\theta | \y_{1:T}) \right].
    \end{equation}
Point estimates, like the MAP, are only simple statistics of the underlying posterior distribution, and fail to signal the overall variability of likely parameters, but it will be useful to use $\hat{\theta}^{\textrm{MAP}}$ when comparing results of estimation from ML and Bayesian techniques. 

From (\ref{eq:MarginalApprox}), it will be sufficient to consider only the prior distribution of parameters and the conditional marginal likelihood distribution (\ref{eq:CondLikelihood}) to form an approximate marginal posterior distribution. Having already defined the marginal likelihood, the remaining factor in estimating the posterior distribution is the choice of prior $p(\theta)$. We discuss some common choices and their properties.

\subsection{Prior Distributions} \label{sec:Priors}

The prior distribution incorporates our beliefs about the nature of the parameters and their structure. These beliefs are not perfect and may be subject to their own uncertainty. The role of a prior distribution is to weakly prefer certain parameter values when no information is present. Upon inspection, the posterior distribution (\ref{eq:FullPosterior}) can be seen as a competition between the likelihood and the prior, where the information from the prior becomes diluted given data. One can also view a prior as a penalty term whose role is to weakly enforce constraints onto the parameters, especially when considering a Gaussian prior and likelihood where the prior can be viewed as a type of Tikhonov regularization. 

There are various labels that have been applied in order to classify priors, see \citet{Gelman2013}, but we will not rely on these terms. We distinguish and delineate different prior densities based on whether they inform scale or sign information and whether the constraint informs certain structures. We will also have cause to highlight the number of so-called 'hyper-parameters' (additional parameters needed in order to fully specify the prior distributions). The presence of a large number of hyper-parameters can increase the flexibility of a prior distribution while also increasing the storage and computational costs of producing random samples from the posterior distribution.

We discuss several options of priors for both $\B$ and $\Q$ of varying degree type that we will perform experiments with. A simple prior would be to assign a normal distribution uniformly to all parameters,
    \begin{equation*}
        \B_{ij} \sim \mathcal{N}(\mu, 1) \qquad \Q_{ij} \sim \mathcal{N}(0, 1),
    \end{equation*}
where $\mathcal{N}(\mu, 1)$ is the Normal distribution with mean $\mu$ and variance $1$. Choosing the normal distribution as a prior does not convey sign information and so it does not enforce $\Q$ to be positive definite, nor does it enforce the eigenvalues of $\B$ to be negative. Instead, this choice of prior is a simple and computationally efficient way to weakly enforce the center of mass of the parameters via location and scale information, i.e. it forces the values of $\B$ to be somewhat close to $\mu$ and values of $\Q$ to be small. Variations to inform sign could be the half-Normal,$\mathcal{N}^+(\mu, \sigma)$, half-Cauchy, Cauchy$^{+}(0, \gamma)$, or $\beta(\alpha, \beta)$.

Consider again the 1-dimensional example problem of estimating $\B$ and $\Q$ from 1000 sampled data $\y$, but now from a Bayesian approach. In \textbf{Fig.~\ref{fig:MLE_MAP_Example}} the priors $\Q \sim \mathcal{N}^{+}(0,1)$ and $\B \sim N(-1, 1)$ are used in conjunction with the marginal likelihood to produce the Bayesian posterior in the second and fourth image respectively. In this example, the truth and Maximum A Posteriori values are calculated. For $\B$, the error in the MAP estimate of roughly 60\% is an improvement over the MLE. While for $\Q$, the MAP estimate is marginally worse than the MLE. 

Using more sophisticated combinations of distributions, we can leverage knowledge about the structures of the matrix parameters in the LIM. Decompose $\Q$ as
    \begin{equation}\label{eq:Qdecompose}
        \Q = \sigma L L^T \sigma,
    \end{equation}
where $\sigma$ is a diagonal matrix with $\sigma_{ii} = \sqrt{Q_{ii}}$ and $L$, a lower triangular matrix, is the Cholesky decomposition of the remaining correlation matrix. We impose simple half-Normal or half-Cauchy prior distributions on the diagonal elements of $\sigma$. For the $m(m+1)/2$ elements of $L$, a popular choice in the literature is to place a LKJ prior on $LL^T$ \citep{Lewandowski2009}, whose distribution is uniform over the space of correlation matrices.  

While mathematical constraints on the spectrum of $\B$ pose a possible source of prior information, placing priors directly on an eigenvalue decomposition of $\B$ is difficult because it would require specification of probabilities in the complex plane and increases the number of parameters needed to be estimated. Furthermore, the eigenvalue decomposition is sensitive to random perturbations in the input matrix to the order of the condition number of the matrix times the perturbation \citep{Stoer2002a} - enough to possibly misidentify elements of the spectrum of $B$. Instead, we use distributions that indirectly produce these qualities by inducing sparsity (only few non-zero elements) and a (nearly) diagonal matrix. By inducing sparsity, we are hypothesizing that EOFs are weekly coupled and this coupling decays with the explained variance of the EOF. We focus on two types of distributions that have been used to induce this kind of sparsity to varying degrees, the Minnesota (Litterman) and (Finnish) Horseshoe priors (see \textbf{Appendix A}). The extent to which sparsity is induced is data-driven and determined by hyper-parameter values, meaning that we do not impose a strict prior belief of sparse interaction between EOFs. 

The Minnesota-Litterman prior \citep{Litterman1986a, Doan1983} prior expresses the belief that (\ref{eq:LSDEsolution}) should be approximately an uncoupled random walk through each dimension. This prior places unit normal priors on the diagonal of $\G(\tau)$ and zero normal priors on the off diagonal components. The variance of these off-diagonal components is then proportional to the variance of the underlying dynamics. This mechanism has the effect of pushing an off-diagonal element to zero if there is not much interaction between the corresponding state variables. There are a total of two hyper-parameters that are used to also control the over-all level of sparsity in the matrix, making this a computationally attractive prior. 

Similar to the Minnesota-Litterman prior, the horseshoe prior \citep{Carvalho2009} was developed to allow for greater control of matrix sparsity. With the horseshoe prior, all entries of $\B$ are given a centered normal distribution whose variance is controlled by a global and local sparsity hyper-parameter. The global hyper-parameter sets a certain variance level for all elements of the matrix, while each entry has its own local hyper-parameter. Therefore, using a horseshoe prior for $\B$, say, requires the learning and sampling of $1+2m^2$ total parameters. The horseshoe formulation, while increasing the storage cost and producing a more topologically complicated posterior distribution to sample from, can, in principle, produce a more nuanced sparsity pattern than the Minnesota-Litterman prior. In practice, a regularized, or Finnish horseshoe prior is used.

Once an adequate prior has been chosen, all that remains is to sample the posterior distribution  (\ref{eq:Marginal}) or the approximation (\ref{eq:MarginalApprox}). There are several methods to achieve this, but in this paper we use a variant of the Markov Chain Monte Carlo method (MCMC)\citep{Brooks2011}, Hamiltonian Monte Carlo (HMC)\citep{Neal2012}, see the \textbf{Appendix B} for details of the theory and implementation. This method is chosen, in particular, because the need for an algorithm scalable to large parameter spaces. HMC utilizes gradient information to produce samples from the target distribution quicker than traditional MCMC methods, reducing the storage and computational costs required. 
    
\section{Parameter Estimation Results}
\label{PER}
We aim to present a test application to demonstrate that the advantages of enhanced estimation and forecast calibration gained from the Bayesian approach is significant. We compare a suite of estimation techniques, including the ML and various MAP estimators corresponding to combinations of the various aforementioned priors. Furthermore, we test the forecast ability using probabilistic metrics to understand the ability of each method to capture the proper forecast distribution. 

We consider the task of forecasting yearly SST anomalies. We use SST data from the fully-coupled 1800 year pre-industrial control run from the CESM1 Large Ensemble project \citep{EKaya}, which are averaged to annual values to produce a time series of 1800 data points for each surface ocean location on a roughly $1^{\circ}$ global grid. The control run is used here in order to test the impact of data availability on parameter estimation, not necessarily to comment on the associated dynamics from forecasting, and to avoid the effects of uncertainty in atmospheric forcing. This study is intended to showcase the potential benefits from a Bayesian approach by directly comparing it with the classical ML methodology in a fair manner. Further research is necessary to come to a comprehensive conclusion about the benefits of the Bayesian approach on realistic data sets. An Empirical Orthogonal Function (EOF) decomposition is performed to reduce the more than 86,000 spatial points to 10 global EOF spatial patterns that capture approximately 55\% of the explained variance. Let $\lbrace \x_t \rbrace$ represent the time series representing the dynamics of the 10 spatial patterns over the 1800 year time frame. A roughly 900 year fraction of this is taken as training data, 500 years is reserved as test data with the remaining portion saved for validation purposes. The matrices $\B, \Q$ require $10 \times 10 = 100$ elements each to be estimated; i.e. the ratio of data points to parameters is 10:1. Increasing the number of EOF spatial patterns reduces this ratio while only obtaining marginal increases in explained variance. We focus our attention on the predictability of the high-latitude regions, and the evidence we display will be concerned with each method's performance in these regions. We believe that using global EOFs may capture possible teleconnections between regions, but recognize that strictly higher-latitude EOFs may capture regional variability using fewer modes.

Here we aim to compare the estimation capabilities of the ML method and the Bayesian approach with a suite of prior distributions. This study performs two experiments: first, a perfect model experiment in order to determine the accuracy of ML and MAP estimates of parameters, and, second, a forecasting experiment in order to test the calibration of the various parameter densities. 

\subsection{Perfect Model Methodology}
\label{PMM}
In this perfect model scenario, we test the ability of each estimation technique to accurately estimate the mean parameter values given data. Using the truncated EOFs, we first estimate MAP values for the parameters $\B$ and $\Q$ using HMC (see the appendix). Using these MAP values, we discretize (\ref{eq:LSDE}) using the standard Euler-Maruyama numerical method and simulate a path with integration time step $\Delta t = 10^{-3}$ years. To sub-sample from this path, we take observations at intervals $\tau = k \Delta t = 1$,  to construct annual time series of length $N = 50,100,1000$. With these time series as data, the mean likelihood or posterior estimates are constructed and compared against the 'true' parameter values for $\B$ and $\Q$. The benefit of using data-inspired parameter values, as opposed to specifying values arbitrarily (e.g. diagonal matrices), is to not pre-impose prior beliefs on the test data or true parameter values and potentially incur bias in the test results. Because of the randomness in the construction of the sample paths and Markov chain construction, the use of HMC estimated parameters as truth should not introduce bias results towards this method.

\begin{figure*}
    \centering
    \includegraphics[width=39pc]{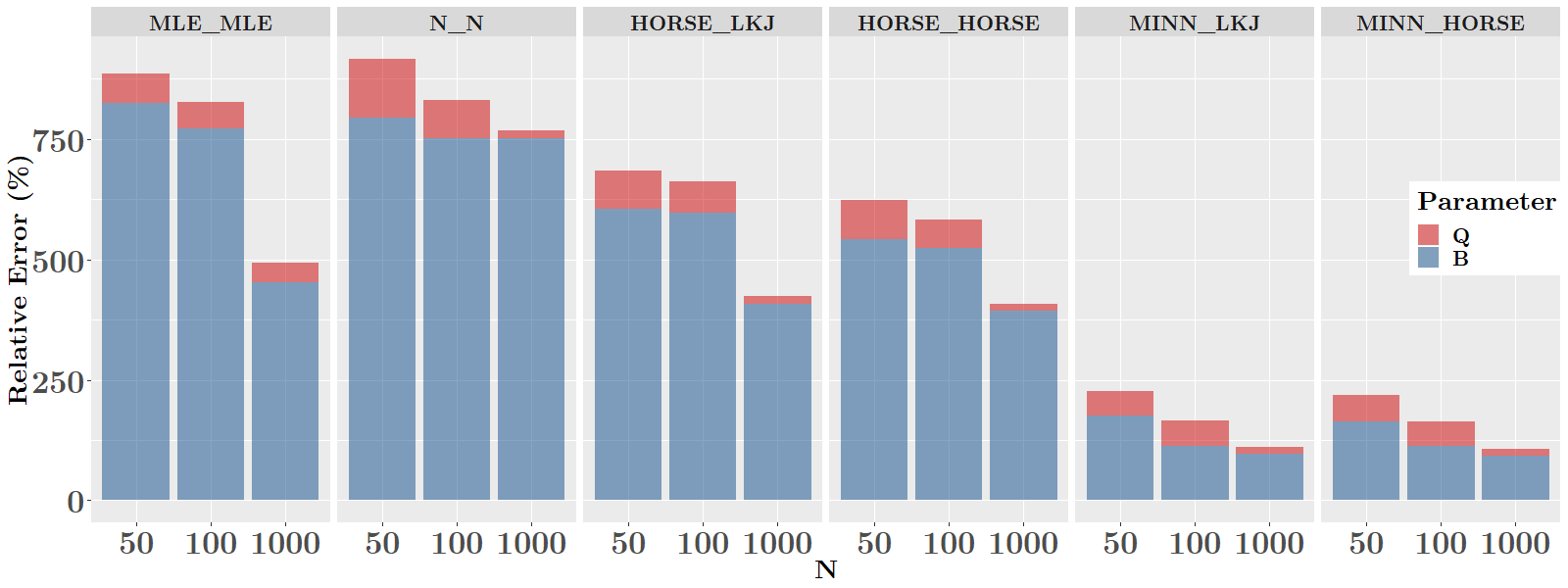}
    \caption{$\|\cdot\|_2 $ Relative errors (\%) for $\B$ and $\Q$ using a variety of prior densities and estimation techniques from the perfect model experiment. Data was generated from an Euler-Maruyama discretization ($\Delta t = 10^{-3}$) of a linear SDE with prescribed drift and diffusion coefficients $\B$ and $\Q$. From this series, $N$ points were uniformly observed for the estimation data set. Relative errors are taken with respect to the ML or MAP estimates of the parameters. We use the notation PriorB\_PriorQ, with 'ML', 'N', 'MINN', 'LKJ', 'HORSE' referring to Maximum Likelihood, Normal, Minnesota, LKJ, and Horseshoe priors respectively.} \label{fig:PerfectModel}
\end{figure*}

For each of the prior distributions that we have discussed, we measure the relative errors (as a percent) of point estimates, $\argmax{p(\theta | y_{1:T})}$, from the 'true' parameter values and results  are shown in \textbf{Fig.~\ref{fig:PerfectModel}}. There are three main conclusions from this experiment. First, as should be no surprise, the relative errors decrease, although somewhat unevenly, as more data is available. The lack of clear convergence, as a result of the `limited' data, should serve as a precaution to those relying on the Central Limit Theorem for asymptotic guarantees. Second, the Bayesian methodology can produce estimates 4-6$\times$ more accurate than ML alone. The choice of Minnesota prior for $\B$ and either LKJ or Horseshoe for $\Q$ produced nearly identical results and features slightly less than 100\% relative error in the large data scenario. Finally, imposing structure in a concise manner is a critical component when considering a prior distribution. There are obviously some prior combinations that perform just as poorly as MLE, such as the normal priors, but priors with many hyper-parameters (like using the Horseshoe prior for $\B$) can also be underwhelming.  

The errors in this experiment may seem unreasonably high to an observer given the relative success of applying LIMs to similar applications. We address this concern in two ways. First, the signal in the yearly averaged CESM control run data is likely much weaker than those used in other papers, such as those from regional or monthly data sets that include atmospheric forcing. The stronger the mean of the signal compared to the variance, the better the parameter estimation results will be. There are also some artificial/ad hoc ways to reduce noise in the data to achieve the same ends, such as smoothing, that we do not perform in this analysis. It should also be noted that while statistical bootstrapping techniques are common ways to improve estimation by inflating the sample size, we did not find discernible advantages of performing such re-sampling and choose not to report the details in this paper. Second, as we will see in the next experiment, an accurate ML or MAP estimate is not necessary for producing reasonable forecasts. While this experiment measures the error in the mean of the posterior distribution, it does not comment on the overall calibration of the probability distributions or how their uncertainties are propagated through to state forecasts. Therefore, it is not necessary contradictory in practice for there to be both large relative errors in parameter recovery while forecast errors are relatively low.

\subsection{CESM1 Data Experiments}
\label{CESM-LE}
When using a LIM, there is traditionally only one source of uncertainty when computing sample paths of the SDE: the noise process with covariance $\Q$. In the Bayesian framework there is an additional source of uncertainty: the distribution of the parameters $\B$ and $\Q$ themselves. Therefore, we hypothesize that forecasts computed with the MLE alone are too conservative and mis-estimate the variance and resulting distribution of the underlying physical process. The goal of this section is to measure the calibration, i.e. how well the method reproduces the true forecast probability distribution, of various sample posterior distributions $p(\x_t|\B, \Q, \y_{t-1})$. 

To perform each experiment we calculate a 100-member ensemble of  with lead time $tau$ using parameters drawn from their respective posterior distribution (in the ML case this is taken to be only the point estimate). The posterior distributions are estimated from the training set of 900 years. Forecasts are initialized from the test data set, independent of the training data set. For this test, the data sets are smoothed with a 3 year backward-looking moving average. For the Bayesian estimates, a value for $\B$ and $\Q$ are drawn anew for each ensemble member for each one-step prediction. From this ensemble, we calculate an array of statistics and metrics to judge the induced accuracy and calibration from each choice of prior distribution. Not only will we compute the correlation coefficient, an oft-used statistic used in time-series forecasting, but we will also calculate the empirical distribution function for the posterior distribution of sample paths and compare to the distribution of the original data using probabilistic metrics in order to have a holistic view of the ability of each method to capture the underlying uncertainty. 

\begin{table}[b]
\begin{center}
\begin{tabular}{ c  c  c  c  c  c }
   \multicolumn{6}{c}{ \qquad Priors for $\Q$} \\ \topline
  $\;$  &   & ML & N & LKJ & HORSE \\ \cline{2-6}
  \multirow{4}{*}{\rotatebox[origin=c]{90}{Priors for $\B$}} \qquad    & ML & \textbf{0.429} & - & - & - \\ \cline{2-6}
  & N & - & 0.852 & 0.852 & 0.851  \\ \cline{2-6}
  & HORSE & - & 0.852 & 0.850  & 0.851 \\ \cline{2-6}
  & MINN & -  & 0.851 & 0.852  & \textbf{0.852} \\ \botline
 \end{tabular}
 \end{center}
 \caption{Correlation Coefficient between the CESM-LE data and the median path from $p(\x_t | \y_t, \B, \Q)$ using a variety of prior densities and estimation techniques. Ensembles of predictions were generated from an Euler-Maruyama discretization ($dt = 10^{-2}$ between data points) of a linear SDE with drift and diffusion coefficients $\B$ and $\Q$ drawn from the associated posterior distribution for each member of the ensemble. From the ensemble the median path was calculated and compared to the original data set. Max Likelihood means that Bayesian estimation was not performed for that variable and only the Maximum Likelihood value was used. }\label{tCESMCorrelation}
\end{table}

\begin{figure*}
    \noindent\includegraphics[width=39pc]{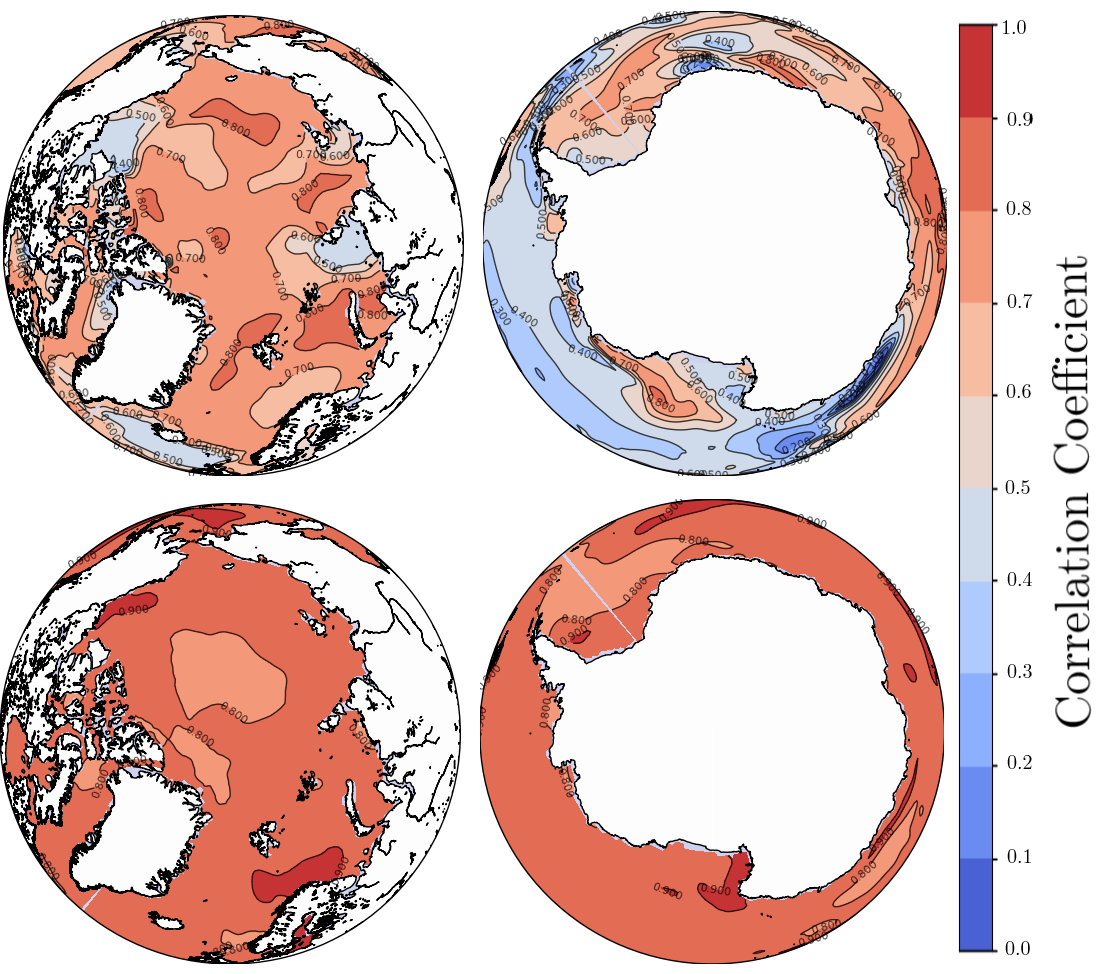}
    \caption{ Correlation Coefficient between mean forecasts with 1 year lead time and test data; higher correlation coefficients are more predictive. (Top) ML estimates were used for the parameter values of $\B$ and $\Q$. (Bottom) Minnesota and LKJ priors were used for $p(\B)$ and $p(\Q)$ respectively. }
    \label{fig:Correlation1year}
\end{figure*}

\begin{figure*}
    \noindent\includegraphics[width=39pc]{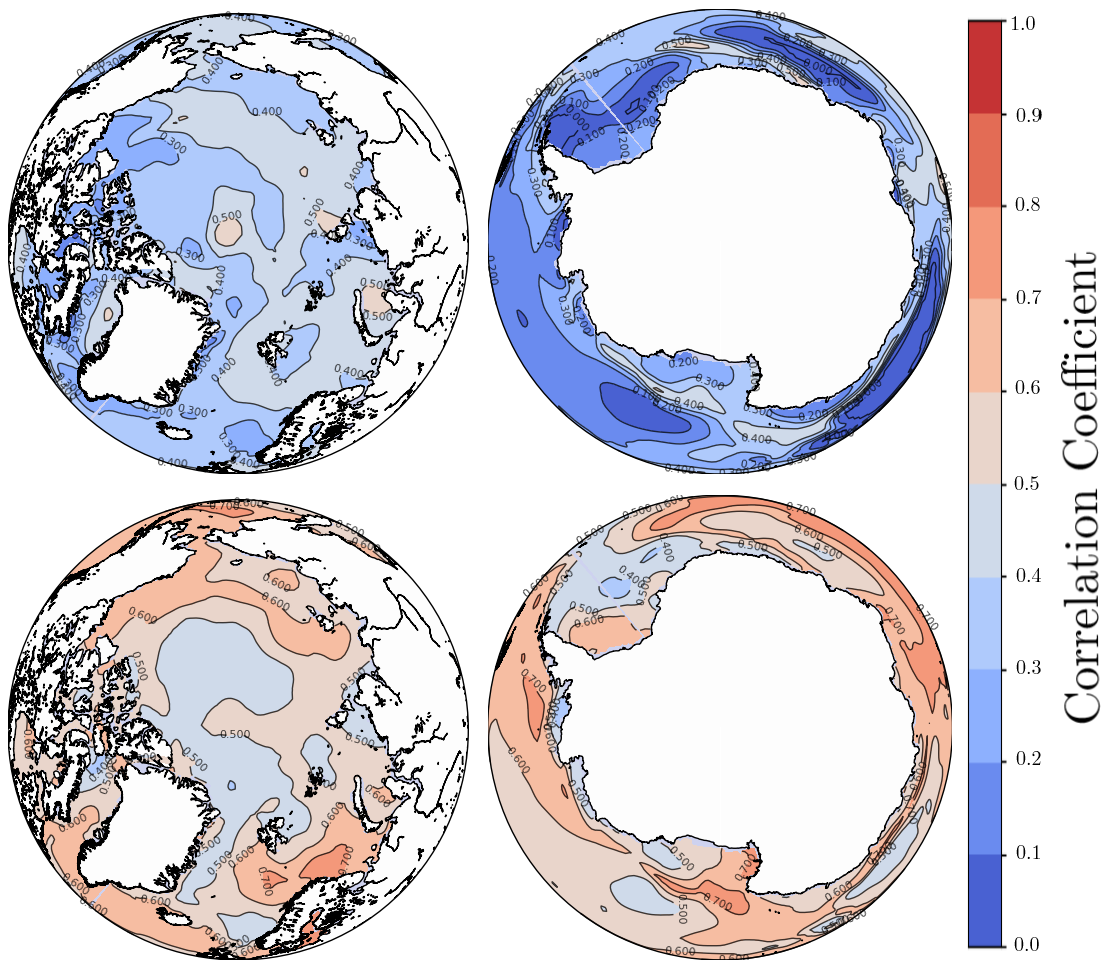}
    \caption{ Correlation Coefficient between mean forecasts with 2 year lead time and test data; higher correlation coefficients are more predictive. (Top) ML estimates were used for the parameter values of $\B$ and $\Q$. (Bottom) Minnesota and LKJ priors were used for $p(\B)$ and $p(\Q)$ respectively. }
    \label{fig:Correlation2year}
\end{figure*}

\textbf{Table~\ref{tCESMCorrelation}} displays the correlation coefficient between the median sample path and test data. In practice a correlation coefficient greater than .6 represents a heuristic for skillful forecasts \citep{Collins2002}. All fully Bayesian forecasting methods exceed this threshold and Hybrid methods with normal and horseshoe priors for $\B$ also produce skillful forecast that rival the fully Bayesian forecasts. Only using a MLE for $\B$ produces poor forecast skill. These two facts together suggest that under-specification of $p(\B)$, not $p(\Q)$, is one of the major causes of poor forecast performance. The spatial, graphical, results for ML and Bayesian estimation using Minnesota and LKJ priors can be seen in \textbf{Fig.~\ref{fig:Correlation1year}} and \textbf{Fig.~\ref{fig:Correlation2year}} for 1 and 2 year lead times, respectively. For forecasts with 1 year lead time, the ML estimates are generally skillful in the Arctic Ocean with exceptions in the Nares strait and Beaufort and Kara seas. In the Southern Ocean, the ML estimates are only skillful off of East Antarctica and in pockets in the Ross and Weddell seas. The Bayesian forecasts has exceptional skill in both polar regions. For forecasts with 2 year lead time, the overall skill of either technique is markedly lower. The ML estimates are not skillful in any region. The Bayesian forecasts are considered skillful still in many areas of both polar regions. In particular we see high correlation skill off of West Antarctica in the Amundsen-Bellingshausen seas, whose climate is known to be strongly influenced by tropical Pacific SSTs \citep{steig2012tropical, ding2011winter, lachlan2006teleconnections}. The correlation metric, then, is very sensitive to proper accounting of the uncertainty in $\B$, but this is not a comprehensive view of the calibration of the forecasts.

\begin{figure*}
    \noindent\includegraphics[width=39pc]{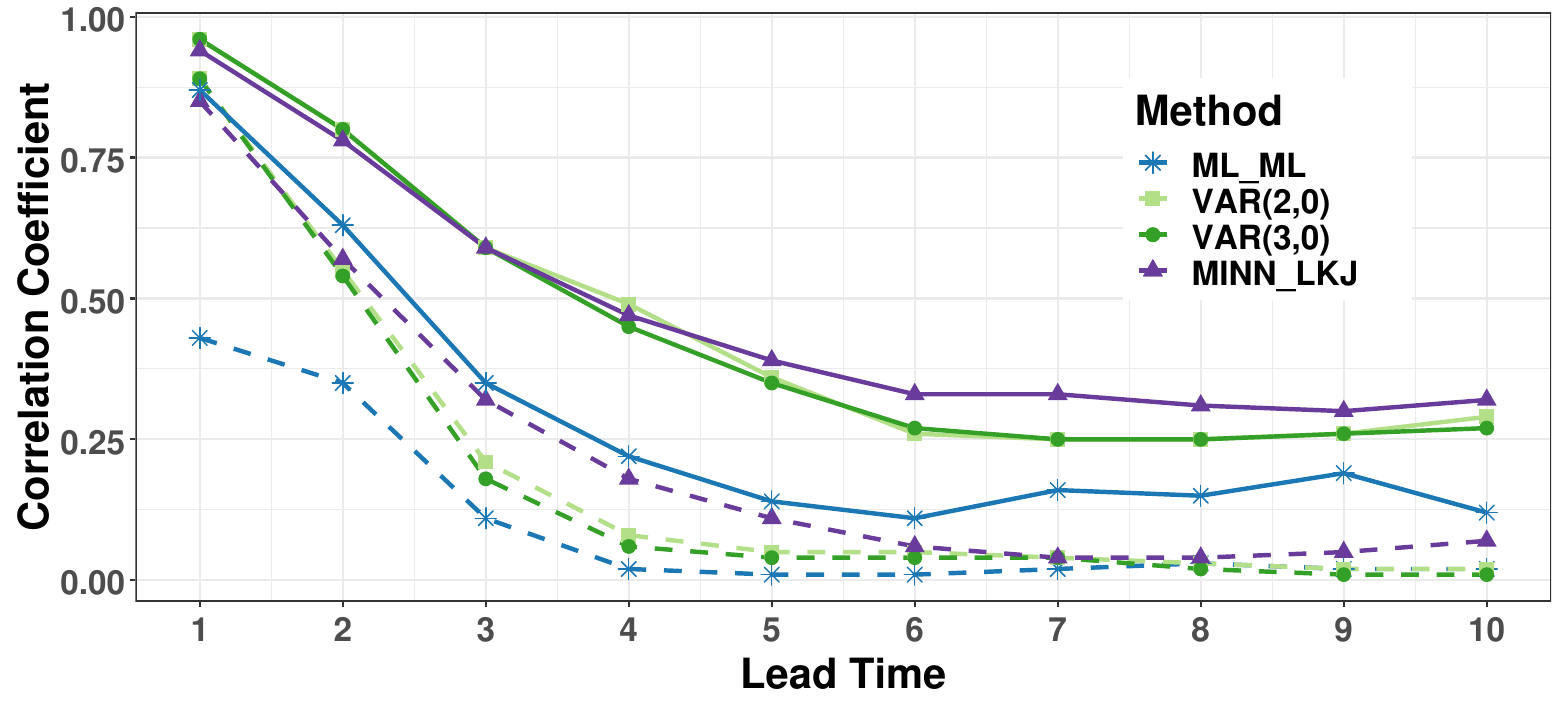}
    \caption{Global mean (dashed) and maximum (solid) correlation coefficients of forecasts from using a variety of methods as a function of the lead (lag) time $\tau$. }
    \label{fig:DecadalCorrelationCoefficient}
\end{figure*}

A view of the correlation coefficient of various methods as a function of lead time is given in \textbf{Fig.~\ref{fig:DecadalCorrelationCoefficient}}. Here, we plot both the global mean (dashed) and global max (solid) correlation coefficients of the ML forecasts, Bayesian forecasts with Minnesota and LKJ priors, and Vector AutoRegressive (VAR) forecasts with 2 and 3 terms. These VAR models are fit with a optimization-based maximum likelihood approach, and are included as a means of comparison of LIMs against other typical time-series forecasting techniques. In general, the ML LIM forecasts are much less skillful than either the Bayesian LIM or VAR models, with the latter models having roughly identical skill. In particular, the low global correlation coefficients and sharp decline of skill at small lead times from the ML forecasts are extreme in comparison to the other methods. Notice that while the ML forecasts have skillful globally averaged correlation coefficients, the low quality of the average correlation coefficients indicates that there is much greater variance of region dependent skill when compared to the other methods, especially for short lead times. All methods, however, produce forecasts that are not skillful on average with a lead time of greater than 3 years. 

\begin{figure*}
    \noindent\includegraphics[width=39pc]{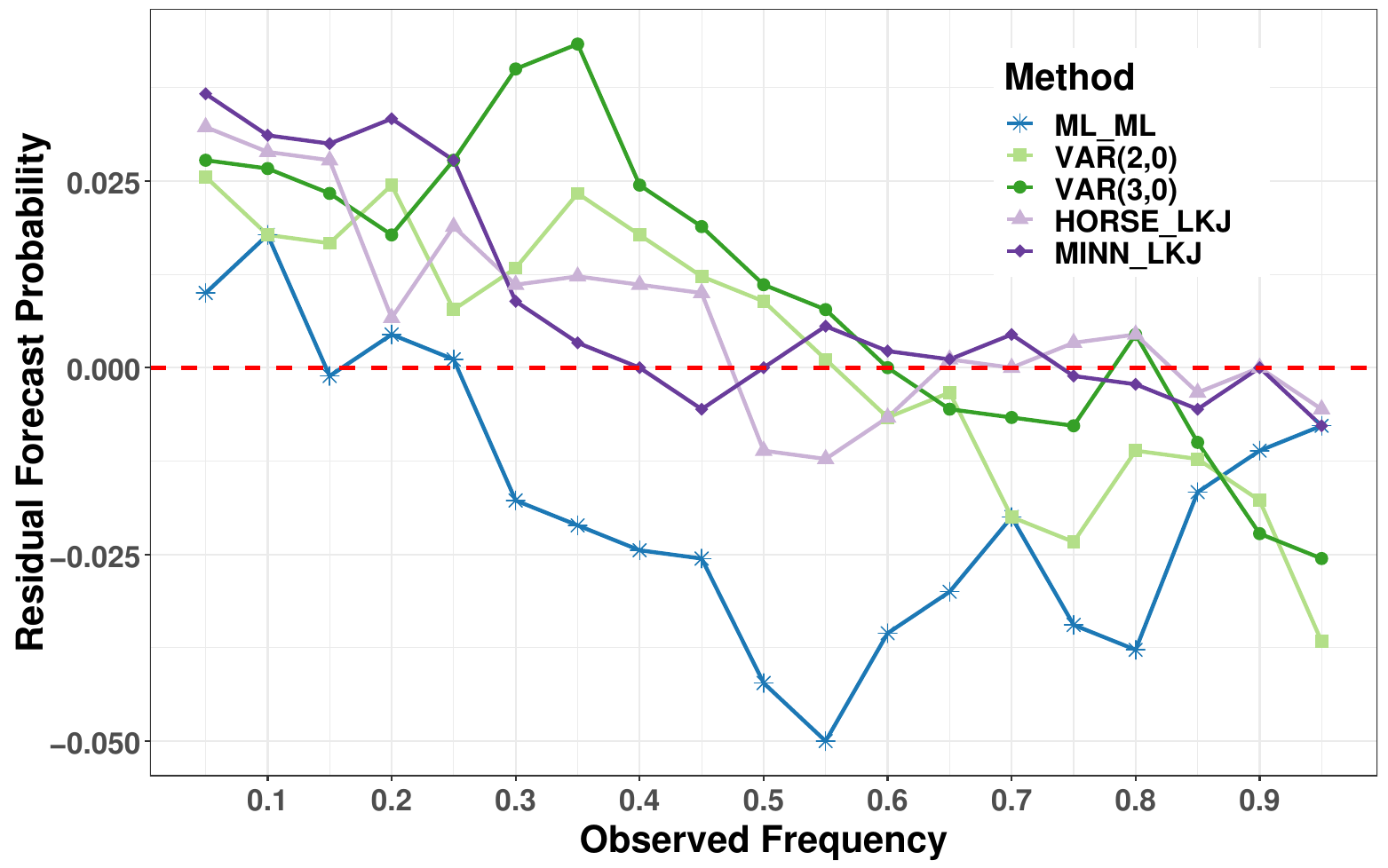}
    \caption{Comparison of empirical distribution function of observed data against ensembles of forecasts using different prior probabilities as well as ML estimated \textrm{VAR}(2,0) and \textrm{VAR}(3,0) methods. The horizontal axis represents the SST anomalies that occur at an observed frequency based on the data and the vertical axis represents the difference between the forecasted and observed frequency of those values. The red dashed line represents an ideal 1-1 correspondence between forecast and observation distribution, positive/negative residuals represent overestimation/underestimation of events of that observed frequency. } \label{fig:CESMDensity}
\end{figure*}

To begin to measure the calibration of probabilistic forecasts, we use the ensemble forecast members to construct conditional cumulative distribution functions, from which quantiles can be calculated and compared to the empirical distribution function for the observations, 
\begin{equation}\label{eq:EDF}
    \hat{F}(t) = \frac{1}{T}\sum_{n=1+\tau}^{T} \pmb{1}_{ (y_{n}-y_{n-\tau}) < t}.
\end{equation}
To be clear, under the assumption of stationarity, this average is performed temporally but for clarity in presentation we also integrate spatially to consider a 1-dimensional distribution function.  \textbf{Fig.~\ref{fig:CESMDensity}} plots the observed and forecast distribution functions against each other, where the discrepancy between forecasted and observed frequencies are plotted. A perfect method would have 1:1 agreement between the forecasts and the observations and have zero residual forecast probability, as denoted in the plot by the dashed (red) line. The distribution of the ML forecasts (dashed) are generally below the dotted line, especially for more frequent events, representing that the method underestimates the uncertainty in the data. The forecasts generated with a Minnesota prior for $\B$ and a Horseshoe prior for $\Q$ have generally good agreement with the observations, but slightly overestimate the uncertainty in unlikely events. We include Vector Auto-Regressive models for a wider comparison of methods, these VAR models are fitted with an optimization based maximum likelihood model. The VAR models suffer more strongly from over- and under-estimation than the Bayesian methods. All of the Bayesian methods outperform the ML techniques in this metric, showing again that regularizing using reasonable prior densities and having access to the entire probability distribution produces more skillful forecasts.

\begin{figure*}
    \noindent\includegraphics[width=39pc]{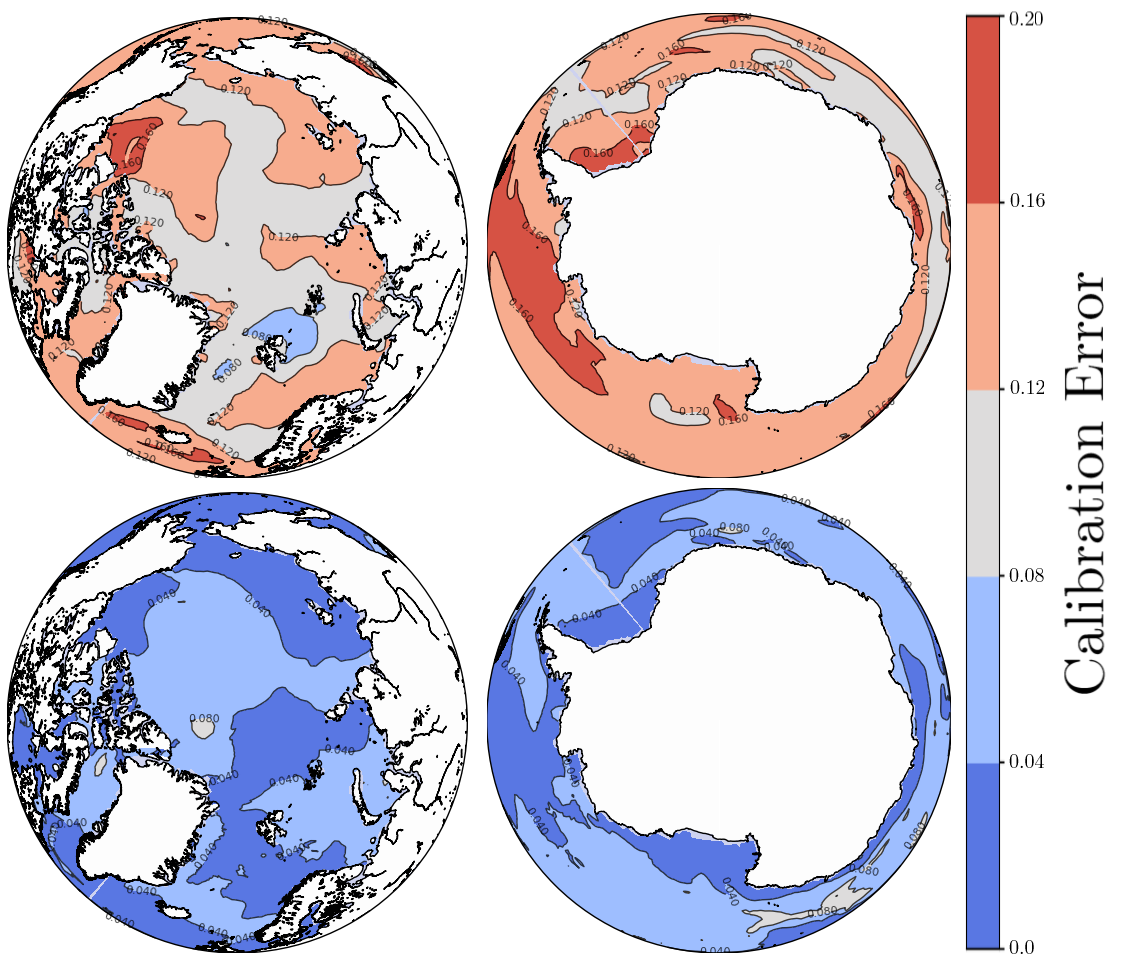}
    \caption{Calibration errors that represent the relative difference between forecast cumulative distribution and the empirical observation distribution. Smaller errors represents a better calibrated forecast. (Top) ML estimates were used for parameter values. (Bottom) Minnesota and LKJ priors were used for $p(\B)$ and $p(\Q)$ respectively. }
    \label{fig:CalibrationScore}
\end{figure*}

\textbf{Fig.~\ref{fig:CalibrationScore}} presents the relative $L_2$ error between the distribution of forecasts at a lead time of 1 year and of observations, what we call the "Calibration Error", as a function of space. Here a lower Calibration Error represents a smaller percent discrepancy between the forecast probability distribution from the empirical observational distribution. Generally, the ML forecasts are at least 1.5 to 2 times as less calibrated than the Bayesian forecasts. 

\section{Conclusion}
\label{CONCLUDE}

In this paper we advocate for a Bayesian approach to LIMs as a way to 1) address the ill-conditioned nature of high-dimensional parameter estimation by providing a formal way to regularize the parameter distributions and 2) produce well-calibrated predictive uncertainties. This strategy has been tested and compared to traditional LIM modeling at recovering LIM parameter values and at forecasting high-latitude SST anomalies. The evidence presented in this paper suggests that a Bayesian approach to statistical forecasting may produce significant recovery and calibration benefits, especially in terms of regional predictability, when compared to simple point-estimation. 

Bayesian approaches to parameter estimation for stochastic differential equations are rare in the literature for a number of factors: there is a lack of closed form solutions to key probability densities, prohibitive computational cost make estimation difficult, and the decent historic performance of various levels of simplifying approximations have not spurred the need to apply the Bayesian approach. Given the increasing amount of computational resources now commonly available, development of efficient sampling algorithms, and evolving view of statistical estimation have set a fertile field for the adoption of the Bayesian framework. This paper advocates for the intelligent use of a Bayesian framework with nontrivial prior distributions to formulate the LIM problem and provides an accounting of how well these approaches reconstruct and forecast the appropriate probability distributions. 

Maximum Likelihood and least square approaches to parameter estimation in LIMs have asymptotic convergence guarantees, but may not be optimal when the available data is sparse, the underlying signal is not strong, there are confounding variables present, or when confronted with model error. As demonstrated by the simple example in \textbf{Fig.~\ref{fig:MLE_MAP_Example}} and \textbf{Fig.~\ref{fig:PerfectModel}}, researchers should be cautious about using such estimators in the LIM framework. We found that, even in a relatively large data set, relative errors often exceeded 100\%. Furthermore, our results also suggest that it would be hazardous to conclude that forecast skill is a reflection of parameter reconstruction \textbf{Table~\ref{tCESMCorrelation}}. 

The main hypothesis of this paper is that using ML point estimates can result in overly conservative forecasts, especially when compared to results from a Bayesian framework. Using both traditional and probabilistic forecast metrics, we found that predictions from a Bayesian approach are more accurate than their ML competitors. Comparing the empirical distribution functions from forecasts and the metrics in \textbf{Table~\ref{tCESMCorrelation}}, \textbf{Fig.~\ref{fig:CESMDensity}}, and \textbf{Fig.~\ref{fig:CalibrationScore}} affirms that the Bayesian approach better captures the probability distribution of the dynamics and makes clear that the benefits from using a Bayesian framework in forecasting can be significant. 

There are several possible avenues to continue and extend the analysis in this paper. Immediately, this analysis should be extended to observational data sets so that comparisons can start to be used with other papers, e.g. \citet{Penland1995}. It is still an open question of whether a Bayesian framework can be used to reliably improve the performance of LIMs in forecasting decadal predictability of SSTs \citep{Hawkins2009, Branstator2010, Guemas2012}, and this area can be pursued with either a global or regional scope. Second, further exploration of relevant prior distributions should be pursued with the goal of producing physically meaningful prior densities. In particular, we can explore the use of global climate models in the Bayesian framework. There are several extensions to LIMs where one could feasibly implement a Bayesian approach. First, investigation of LIM frameworks that incorporate state-depent noise, as in \citet{Sardeshmukh2015} and \citet{MartinezVillalobos2019}, can also be formulated in a Bayesian way given an appropriate approximate likelihood distribution. Finally, one could depart from the parametric form that a linear stochastic differential equation imposes and attempt a Bayesian framework with more complicated, non-parametric physically constrained statistical models such as in \citet{Majda} \citet{Wikle2011} or machine learning techniques such as Bayesian networks in \citet{McDermott2019} and \citet{McDermott2019a}. 
%
\acknowledgments
This research was supported in part by the Regional and Global Model Analysis (RGMA) component of the Earth and Environmental System Modeling (EESM) program of the U.S. Department of Energy's Office of Science, as contribution to the HiLAT-RASM project. DF, NMU, and DC received support for this project from the U.S. Department of Energy, Biological and Environmental Research (BER) division (Regional and Global Modeling and Applications program); NMU received additional support under the BER Early Career Research program.

%






%
%
%
\appendix[A]
\label{appendixA}
\appendixtitle{Prior Matrix Distributions}
\subsection{LKJ}
The so-called LKJ prior \citep{Lewandowski2009} (named after authors: Lewandowski, Kurowicka, and Joe) is the result of work that extended methods designed to sample random correlation matrices in a computationally efficient manner. In implementation, there is a single hyper-parameter to the distribution, $\eta$, and the distribution can be defined as
    \begin{equation} \label{eq:LKJdistribution}
        \textrm{LKJ}(\Sigma | \eta ) \propto \det{\left(\Sigma\right)}^{\eta-1}.    
    \end{equation}
From the definition, if $\eta = 1$, the LKJ distribution is uniform over all correlation matrices. For $\eta > 1$, the distribution places mass closer to $1$, resulting in samples close to the unit matrix. In other words, $\eta$ controls the expected amount of correlation, with $\eta>1$ favoring less correlation, while $\eta<1$ expecting more correlation.

\subsection{Minnesota-Litterman}
The Minnesota-Litterman prior \citep{Litterman1986a, Doan1983} has two hyper-parameters, $\lambda$ and $\theta$, that allow for controlling the degree of sparsity in the matrix and differentiating between diagonal and off-diagonal terms. The prior distributions are given by
    \begin{equation}\label{eq:minnesota}
    \begin{aligned}
        \G_{ii} &\sim N(1, \lambda), \\
        \G_{ij} &\sim N\left(0, \lambda \theta \frac{\Sig_{ii}}{\Sig_{ij}}\right) \quad i\neq j, \\
        \lambda &\sim \text{Cauchy}^{+}(0, 1) \\
        \theta &\sim \mathcal{U}(0, 1), 
    \end{aligned}
    \end{equation}
where $\mathcal{U}(a,b)$ is the uniform distribution on $[a,b]$.

This formulation allows for the distinct treatment of diagonal from off-diagonal parameters. $\lambda$ acts as a global covariance hyper-parameter, controlling the overall uncertainty in the elements of the matrix. If $\lambda = 0$, then the prior would collapse to a distribution around the identity matrix. For $\lambda \neq 0$, the off-diagonal terms $\lambda \theta \Sig_{ii}/\Sig_{jj}$ shrinks $\lambda$ by a multiple, $\theta$, of the relative size of the respective variances. This combination is a natural choice since it controls the sparsity of the matrix according to the relative uncertainty and size of the dynamics of elements of $\x_t$.

\subsection{(Finnish) Horseshoe}
Similar to the Minnesota-Litterman prior, the horseshoe prior \citep{Carvalho2009} was originally derived to allow for specification of sparsity and has hyper-parameters $\lambda$ and $\tau$. $\lambda \in \R^{m \times m}$, however, has $m^2$ elements $\lambda_{ij}$ - one for each entry of $\B$. The distributions are
    \begin{equation}\label{eq:horseshoe}
        \begin{aligned}
            \B_{ij} &\sim N(0, \tau^2 \lambda^2_{ij}), \\
        \lambda_{ij} &\sim \text{Cauchy}^{+}(0, 1),
        \end{aligned}
    \end{equation}
where $\tau>0$ is some chosen constant. The two hyper-parameters $\tau$ and $\lambda$ are interpreted at global and local sparsity controls. The value of $\tau$ should represent our overall belief about the size of elements of $\B$. Then, each $\lambda_{ij}$ is used to tune the size of individual elements $\B_{ij}$. $\lambda_{ij}$ is given a half-Cauchy prior in order to allow for a large spread of values. If a value of $\tau$ is not known a priori, then one can use a non-informative prior like a half-Cauchy instead. Of concern with this scheme is the introduction of $m^2+1$ hyper-parameters that must be estimated along with the $n^2$ elements of $\B$. While the large number of hyper-parameters can help control the size of individual elements $\B_{ij}$, it increases the computational complexity of estimation and may require relatively more data to get reliable estimates than simpler priors. Furthermore, it is occasionally possible during estimation for $\tau^2 \lambda^2$ to become exceedingly large. The Finnish horseshoe introduces a transformation, $\bar{\lambda}_{ij}$, of $\lambda_{ij}$ and additional hyper-parameter $c^2$, such that
    \begin{equation}\label{eq:FinnishHorseshoe}
        \bar{\lambda}^2_{ij} = \dfrac{ \lambda_{ij}^2}{1+\tau^2\lambda_{ij}^2/c^2}.
    \end{equation}
 If $\tau^2 \lambda^2 \ll c^2$, then $\bar{\lambda} \approx \lambda$ and the regularized horseshoe behaves similarly to the original horseshoe prior. On the other-hand, if $\tau^2\lambda^2 \gg c^2$ then $\bar{\lambda}\ll \lambda$ and $\bar{\lambda}$ is a regularization. This transformation penalizes values of $\lambda$ and $\tau$ that are near some predetermined threshold $c^2$. Like $\tau$, $c$ can be estimated by considering the understanding of the size of the parameters, or can be given some non-informative prior and estimated along with the other parameters. Unlike the Minnesota prior, this prior can be applied to both $\B$ and $\Q$. Of course, when estimating $\Q$, we follow the transformation (\ref{eq:Qdecompose}) and need only estimate $L$.
 
\appendix[B]
\label{appendixB}
\appendixtitle{Bayesian Computational Techniques}
To make full use of the Bayesian framework, instead of relying on maximal parameter estimates, we must be able to draw samples of the parameters from their respective probability distributions. If the marginal likelihood function $p(\y_{1:T}|\theta)$ is normal and the prior distribution $p(\theta)$ is chosen conjugate to the likelihood, then the posterior distribution can be calculated analytically. There are numerous books written on this technique, see \citet{Gelman2013}. In general, however, numerical techniques are needed to sample from an arbitrary posterior distribution. In particular, Markov Chain Monte Carlo (MCMC) algorithms are standard practice to generate independent samples from the posterior. We will review MCMC in order to motivate Hamiltonian Monte Carlo (HMC), which is the algorithm we use for the studies in this paper. This appendix is only meant as a brief description of the types of methods employed in this paper, for more detailed and comprehensive discussions see \citet{Brooks2011}. 

\subsubsection{Markov Chain Monte Carlo}
A Markov chain is a sequence of random variables $\lbrace \epsilon_1, \epsilon_2, \dots \rbrace$ where 
    \begin{equation}\label{eq:markov}
        p(\epsilon_1, \epsilon_2, \dots, \epsilon_N) = p(\epsilon_1) \prod_{k=2}^{N} p(\epsilon_k | \epsilon_{k-1} ), 
    \end{equation}
i.e. a member of the sequence, given the previous value, is conditionally independent of all other elements in the sequence. The goal of MCMC is to generate a Markov chain such that elements of the chain come from the posterior distribution. The elements of the Markov chain are then the requested samples. The basic MCMC algorithm consists of specifying an initial distribution $p(\epsilon_1)$, proposing a next element using a transition distribution $p(\epsilon_{k}|\epsilon_{k-1})$, and a method for accepting or rejecting the proposed value. In the LIM case, the initial distribution could be specified by the ML distribution since one has analytical representations for the mean and standard deviation. For this method to generate a proper Markov chain that has $p(\theta|\y_{1:T})$ as its equilibrium distribution the transition density and acceptance criterion must meet certain theoretical conditions. If these conditions are met, then $p(\epsilon_k)$ is stationary, $p(\epsilon_k) = p(\epsilon_{k'})$ for all $k, k'$ and $\epsilon_k$ is independent of $\epsilon_{k'}$. This stationary distribution is then guaranteed to be equal to $p(\theta | \y_{1:T})$. In practice, ensuring that generated Markov chains are independent and come from the required distribution is non-trivial. We will not discuss these theoretical concerns, but assume such conditions have been met and refer the reader to the literature. In order to avoid some of these technical difficulties, we rely on using the Hamiltonian Monte Carlo algorithm.

\subsubsection{Hamiltonian Monte Carlo}
Hamiltonian Monte Carlo (HMC) uses the dynamics from an associated Hamiltonian dynamical system to construct a transition density $p(\epsilon_k | \epsilon_{k-1})$ that is theoretically guaranteed to produce the required posterior distribution. These dynamics are defined by the pair of differential equations
    \begin{equation}\label{eq:HamiltonianDynamics}
    \begin{aligned}
        \frac{d\pmb{p}}{dt} &= -\frac{\partial\pmb{H}}{\partial\pmb{q}}, \\
        \frac{d\pmb{q}}{dt} &= \frac{\partial\pmb{H}}{\partial\pmb{p}},
    \end{aligned}
    \end{equation}
where $(\pmb{p}, \pmb{q})$ are coordinates of the state system (representing momentum and position). $\pmb{H}$ is the Hamiltonian defined as 
    \begin{equation}\label{eq:Hamiltonian}
        \pmb{H}(\pmb{p}, \pmb{q}) = \pmb{K}(\pmb{p}, \pmb{q}) + \pmb{U}(\pmb{p}, \pmb{q}),
    \end{equation}
where $\pmb{K}$ is called the "kinetic energy" and $\pmb{U}$ is the "potential energy". HMC utilizes this framework by introducing a conjugate momentum variable $\rho \sim N(0, M)$ and seeking to draw from the joint density $p(\rho, \theta) = p(\rho | \theta) p(\theta|\y_{1:T})$. This joint density defines the Hamiltonian, 
    \begin{equation}\label{eq:HMCHamiltonian}
        \begin{aligned}
        \pmb{H}(\rho, \theta) &= -\log{p(\rho, \theta})\\
        &= -\log{p(\rho|\theta)} - \log{p(\theta|\y_{1:T})}\\
        &= \pmb{K}(\rho, \theta) + \pmb{U}(\theta)
        \end{aligned}
    \end{equation}
We assume that the momentum is independent of the target density, so $p(\rho|\theta) = p(\rho)$ and $\pmb{K}(\rho, \theta) = \pmb{K}(\rho) = \frac{1}{2} p^T M^{-1} p$. To generate transitions for the Markov chain we evolve the system (\ref{eq:HamiltonianDynamics}) starting from an initial value for $\theta$ and a draw $\rho$ from $p(\rho)$. In particular, we solve
    \begin{equation}\label{eq:HamiltonianDynamics2}
    \begin{aligned}
        \frac{d\theta}{dt} &= \frac{\partial\pmb{K}}{\partial\rho}, \\
        \frac{d\rho}{dt} &= -\frac{\partial\pmb{U}}{\partial\theta}.
    \end{aligned}
    \end{equation}
Solving \ref{eq:HamiltonianDynamics2} in general requires discretization of $\rho$ and $\theta$ into their respective Markov chains and numerical integration. While there are numerous numerical integrators for Hamiltonian systems, the most common in practice is the Leapfrog, or St\"{o}mer-Verlet integrator with step size $\Delta t$
    \begin{equation}\label{eq:Leapfrog}
    \begin{aligned}
        \rho_{n+\frac{1}{2}} &= \rho_{n} - \frac{\Delta t}{2} \frac{\partial \pmb{U}}{\partial \theta},\\
        \theta_{n+1} &= \theta_{n} + \Delta t M^{-1}\rho_{n+\frac{1}{2}},\\
        \rho_{n+1} &= \rho_{n+\frac{1}{2}} - \frac{\Delta t}{2} \frac{\partial \pmb{U}}{\partial \theta}.
    \end{aligned}
    \end{equation}
The probability that these values are accepted is given by
    \begin{equation*}
        \min \left(1, \exp{\left( \pmb{H}(\rho_n, \theta_n) - \pmb{H}(\rho_{n+1}, \theta_{n+1})\right)} \right). 
    \end{equation*}
If the generated values are not accepted then $\theta_{n+1} = \theta_{n}$ and the algorithm continues for a total number of $L$ time steps. 

It is proven that this algorithm produces samples that converge to the target distribution given certain conditions. Often this convergence is quicker, in terms of samples needed, than standard MCMC algorithms. As a trade off for speed, the HMC requires calculation of derivatives at each time step. Furthermore, the time step $\Delta t$ may need to be taken excessively small in order to ensure stability of the scheme, meaning that more time steps are required to adequately explore the state space. Modifications to the original HMC algorithm, the no U-turn sampler in particular, help automatically adjust the step size and number of steps. We leave the details of these algorithms to \citep{Neal2012}. For out-of-the-box implementation of this algorithm, see the programming language Stan \citep{Carpenter2017}.

\bibliographystyle{ametsoc2014}
\bibliography{Bayes_SST}

\begin{thebibliography}{78}
\providecommand{\natexlab}[1]{#1}
\providecommand{\url}[1]{\texttt{#1}}
\renewcommand{\UrlFont}{\rmfamily}
\providecommand{\urlprefix}{URL }
\expandafter\ifx\csname urlstyle\endcsname\relax
  \providecommand{\doi}[1]{doi:\discretionary{}{}{}#1}\else
  \providecommand{\doi}{doi:\discretionary{}{}{}\begingroup
  \urlstyle{rm}\Url}\fi
\providecommand{\eprint}[2][]{\url{#2}}

\bibitem[{Alexander et~al.(2008)}]{Alexander2008}
Alexander, M.~A., and Coauthors, 2008: {Forecasting Pacific SSTs: Linear
  Inverse Model Predictions of the PDO}. \textit{Journal of Climate},
  \textbf{21~(2)}, 385--402, \doi{10.1175/2007JCLI1849.1},
  \urlprefix\url{http://journals.ametsoc.org/doi/abs/10.1175/2007JCLI1849.1}.

\bibitem[{Barnston et~al.(1999)Barnston, He, Glantz, Barnston, Glantz,, and
  He}]{Barnston1999}
Barnston, A.~G., Y.~He, M.~H. Glantz, A.~G. Barnston, M.~H. Glantz, and Y.~He,
  1999: {Predictive Skill of Statistical and Dynamical Climate Models in SST
  Forecasts during the 1997—98 El Ni{\~{n}}o Episode and the 1998 La
  Ni{\~{n}}a Onset}. \textit{Bulletin of the American Meteorological Society},
  \textbf{80~(2)}, 217--243,
  \doi{10.1175/1520-0477(1999)080<0217:PSOSAD>2.0.CO;2},
  \urlprefix\url{http://journals.ametsoc.org/doi/abs/10.1175/1520-0477%281999%29080%3C0217%3APSOSAD%3E2.0.CO%3B2}.

\bibitem[{Barnston et~al.(2012)Barnston, Tippett, L'Heureux, Li,, and
  Dewitt}]{Barnston2012a}
Barnston, A.~G., M.~K. Tippett, M.~L. L'Heureux, S.~Li, and D.~G. Dewitt, 2012:
  {Skill of real-time seasonal ENSO model predictions during 2002-11: Is our
  capability increasing?} \textit{Bulletin of the American Meteorological
  Society}, \textbf{93~(5)}, 631--651, \doi{10.1175/BAMS-D-11-00111.1},
  \urlprefix\url{http://journals.ametsoc.org/doi/abs/10.1175/BAMS-D-11-00111.1}.

\bibitem[{Batz et~al.(2018)Batz, Ruttor,, and Opper}]{Batz2017}
Batz, P., A.~Ruttor, and M.~Opper, 2018: {Approximate Bayes learning of
  stochastic differential equations}. Tech. Rep.~2.
  \doi{10.1103/PhysRevE.98.022109},
  \urlprefix\url{https://arxiv.org/pdf/1702.05390.pdf}.

\bibitem[{Beskos et~al.(2006)Beskos, Papaspiliopoulos, Roberts,, and
  Fearnhead}]{Beskos2006}
Beskos, A., O.~Papaspiliopoulos, G.~O. Roberts, and P.~Fearnhead, 2006: {Exact
  and computationally efficient likelihood-based estimation for discretely
  observed diffusion processes}. \textit{Journal of the Royal Statistical
  Society. Series B: Statistical Methodology}, \textbf{68~(3)}, 333--382,
  \doi{10.1111/j.1467-9868.2006.00552.x},
  \urlprefix\url{https://www.jstor.org/stable/3879281}.

\bibitem[{Branstator and Teng(2010)Branstator, and Teng}]{Branstator2010}
Branstator, G., and H.~Teng, 2010: {Two limits of initial-value decadal
  predictability in a CGCM}. \textit{Journal of Climate}, \textbf{23~(23)},
  6292--6311, \doi{10.1175/2010JCLI3678.1},
  \urlprefix\url{http://journals.ametsoc.org/doi/abs/10.1175/2010JCLI3678.1}.

\bibitem[{Brooks(2011)}]{Brooks2011}
Brooks, S., 2011: \textit{{Handbook of Markov chain Monte Carlo}}. CRC
  Press/Taylor {\&} Francis, 592 pp.,
  \urlprefix\url{https://www.crcpress.com/Handbook-of-Markov-Chain-Monte-Carlo/Brooks-Gelman-Jones-Meng/p/book/9781420079418}.

\bibitem[{Carpenter et~al.(2017)}]{Carpenter2017}
Carpenter, B., and Coauthors, 2017: {Stan : A Probabilistic Programming
  Language}. \textit{Journal of Statistical Software}, \textbf{76~(1)}, 1--32,
  \doi{10.18637/jss.v076.i01},
  \urlprefix\url{http://www.jstatsoft.org/v76/i01/}.

\bibitem[{Carvalho et~al.(2009)Carvalho, Polson,, and Scott}]{Carvalho2009}
Carvalho, C.~M., N.~G. Polson, and J.~G. Scott, 2009: {Handling Sparsity via
  the Horseshoe}. \textit{Journal of Machine Learning Research, W{\&}CP 5
  (AIStats).}, \textbf{5}, 73--80,
  \urlprefix\url{http://proceedings.mlr.press/v5/carvalho09a/carvalho09a.pdf}.

\bibitem[{Chen et~al.(2016)Chen, Cane, Henderson, Lee, Chapman, Kondrashov,,
  and Chekroun}]{Chenb}
Chen, C., M.~A. Cane, N.~Henderson, D.~E. Lee, D.~Chapman, D.~Kondrashov, and
  M.~D. Chekroun, 2016: {Diversity, nonlinearity, seasonality, and memory
  effect in ENSO simulation and prediction using empirical model reduction}.
  \textit{Journal of Climate}, \textbf{29~(5)}, 1809--1830,
  \doi{10.1175/JCLI-D-15-0372.1}.

\bibitem[{Collins(2002)}]{Collins2002}
Collins, M., 2002: {Climate predictability on interannual to decadal time
  scales: The initial value problem}. \textit{Climate Dynamics},
  \textbf{19~(8)}, 671--692, \doi{10.1007/s00382-002-0254-8},
  \urlprefix\url{http://link.springer.com/10.1007/s00382-002-0254-8}.

\bibitem[{Colman and Davey(2003)Colman, and Davey}]{Colman2003a}
Colman, A.~W., and M.~K. Davey, 2003: {Statistical prediction of global
  sea-surface temperature anomalies}. \textit{International Journal of
  Climatology}, \textbf{23~(14)}, 1677--1697, \doi{10.1002/joc.956},
  \urlprefix\url{www.interscience.wiley.com}.

\bibitem[{Davis and Davis(1976)Davis, and Davis}]{Davis1976}
Davis, R.~E., and R.~E. Davis, 1976: {Predictability of Sea Surface Temperature
  and Sea Level Pressure Anomalies over the North Pacific Ocean}.
  \textit{Journal of Physical Oceanography}, \textbf{6~(3)}, 249--266,
  \doi{10.1175/1520-0485(1976)006<0249:POSSTA>2.0.CO;2},
  \urlprefix\url{http://journals.ametsoc.org/doi/abs/10.1175/1520-0485%281976%29006%3C0249%3APOSSTA%3E2.0.CO%3B2}.

\bibitem[{DelSole et~al.(2003)DelSole, Chang, DelSole,, and
  Chang}]{DelSole2003}
DelSole, T., P.~Chang, T.~DelSole, and P.~Chang, 2003: {Predictable Component
  Analysis, Canonical Correlation Analysis, and Autoregressive Models}.
  \textit{Journal of the Atmospheric Sciences}, \textbf{60~(2)}, 409--416,
  \doi{10.1175/1520-0469(2003)060<0409:PCACCA>2.0.CO;2},
  \urlprefix\url{http://journals.ametsoc.org/doi/abs/10.1175/1520-0469%282003%29060%3C0409%3APCACCA%3E2.0.CO%3B2}.

\bibitem[{DelSole et~al.(1999)DelSole, Hou, DelSole,, and Hou}]{DelSole1999}
DelSole, T., A.~Y. Hou, T.~DelSole, and A.~Y. Hou, 1999: {Empirical Stochastic
  Models for the Dominant Climate Statistics of a General Circulation Model}.
  \textit{Journal of the Atmospheric Sciences}, \textbf{56~(19)}, 3436--3456,
  \doi{10.1175/1520-0469(1999)056<3436:ESMFTD>2.0.CO;2},
  \urlprefix\url{http://journals.ametsoc.org/doi/abs/10.1175/1520-0469%281999%29056%3C3436%3AESMFTD%3E2.0.CO%3B2}.

\bibitem[{Delsole et~al.(2013)Delsole, Jia,, and Tippett}]{DelSole2013}
Delsole, T., L.~Jia, and M.~K. Tippett, 2013: {Decadal prediction of observed
  and simulated sea surface temperatures}. \textit{Geophysical Research
  Letters}, \textbf{40~(11)}, 2773--2778, \doi{10.1002/grl.50185},
  \urlprefix\url{http://doi.wiley.com/10.1002/grl.50185}.

\bibitem[{Delsole and Yang(2010)Delsole, and Yang}]{DelSole2010}
Delsole, T., and X.~Yang, 2010: {State and parameter estimation in stochastic
  dynamical models}. \textit{Physica D: Nonlinear Phenomena},
  \textbf{239~(18)}, 1781--1788, \doi{10.1016/j.physd.2010.06.001},
  \urlprefix\url{https://www.sciencedirect.com/science/article/pii/S0167278910001673}.

\bibitem[{Dias et~al.(2018)Dias, Subramanian, Zanna,, and Miller}]{Dias2018}
Dias, D.~F., A.~Subramanian, L.~Zanna, and A.~J. Miller, 2018: {Remote and
  local influences in forecasting Pacific SST: a linear inverse model and a
  multimodel ensemble study}. \textit{Climate Dynamics}, 1--19,
  \doi{10.1007/s00382-018-4323-z},
  \urlprefix\url{http://link.springer.com/10.1007/s00382-018-4323-z}.

\bibitem[{DiNezio et~al.(2017)DiNezio, Deser, Okumura,, and
  Karspeck}]{DiNezio2017}
DiNezio, P.~N., C.~Deser, Y.~Okumura, and A.~Karspeck, 2017: {Predictability of
  2-year La Ni{\~{n}}a events in a coupled general circulation model}.
  \textit{Climate Dynamics}, \textbf{49~(11-12)}, 4237--4261,
  \doi{10.1007/s00382-017-3575-3},
  \urlprefix\url{http://link.springer.com/10.1007/s00382-017-3575-3}.

\bibitem[{Ding et~al.(2011)Ding, Steig, Battisti,, and
  K{\"{u}}ttel}]{ding2011winter}
Ding, Q., E.~J. Steig, D.~S. Battisti, and M.~K{\"{u}}ttel, 2011: {Winter
  warming in West Antarctica caused by central tropical Pacific warming}.
  \textit{Nature Geoscience}, \textbf{4~(6)}, 398--403, \doi{10.1038/ngeo1129},
  \urlprefix\url{http://www.nature.com/articles/ngeo1129}.

\bibitem[{Doan et~al.(1983)Doan, Litterman, Sims,, and Sims}]{Doan1983}
Doan, T., R.~Litterman, C.~A. Sims, and C.~Sims, 1983: {NBER WORKING PAPER
  SERIES FORECASTING AND CONDITIONAL PROJECTION USING REALISTIC PRIOR
  DISTRIBUTIONS Forecasting and Conditional Projection Using Realistic Prior
  Distributions}. Tech. rep.
  \urlprefix\url{https://www.nber.org/papers/w1202.pdf}.

\bibitem[{Eraker(2001)}]{Eraker2001}
Eraker, B., 2001: {MCMC analysis of diffusion models with application to
  finance}. \textit{Journal of Business and Economic Statistics},
  \textbf{19~(177-191)},
  \urlprefix\url{http://eraker.marginalq.com/eraker01JBES.pdf}.

\bibitem[{Gelman et~al.(2013)Gelman, Carlin, Stern, Dunson, Vehtari,, and
  Rubin}]{Gelman2013}
Gelman, A., J.~B. Carlin, H.~S. Stern, D.~B. Dunson, A.~Vehtari, and D.~B.
  Rubin, 2013: \textit{{Bayesian data analysis, third edition}}. Chapman {\&}
  Hall/CRC Texts in Statistical Science, Taylor {\&} Francis, 1--646 pp.,
  \urlprefix\url{https://books.google.com/books?id=ZXL6AQAAQBAJ}.

\bibitem[{Giannone et~al.(2015)Giannone, Lenza,, and Primiceri}]{Giannone2015}
Giannone, D., M.~Lenza, and G.~E. Primiceri, 2015: {PRIOR SELECTION FOR VECTOR
  AUTOREGRESSIONS.} \textit{Review of Economics {\&} Statistics},
  \textbf{97~(2)}, 436--451,
  \urlprefix\url{http://faculty.wcas.northwestern.edu/~gep575/Draft_GLP_V24.pdf
  https://search.ebscohost.com/login.aspx?direct=true&db=bth&AN=102240456&site=ehost-live}.

\bibitem[{Guemas et~al.(2012)Guemas, Doblas-Reyes, Lienert, Soufflet,, and
  Du}]{Guemas2012}
Guemas, V., F.~J. Doblas-Reyes, F.~Lienert, Y.~Soufflet, and H.~Du, 2012:
  {Identifying the causes of the poor decadal climate prediction skill over the
  North Pacific}. \textit{Journal of Geophysical Research Atmospheres},
  \textbf{117~(20)}, \doi{10.1029/2012JD018004},
  \urlprefix\url{http://doi.wiley.com/10.1029/2012JD018004}.

\bibitem[{Guemas et~al.(2016)}]{guemas2016review}
Guemas, V., and Coauthors, 2016: {A review on Arctic sea-ice predictability and
  prediction on seasonal to decadal time-scales}. \textit{Quarterly Journal of
  the Royal Meteorological Society}, \textbf{142~(695)}, 546--561,
  \doi{10.1002/qj.2401}, \urlprefix\url{http://doi.wiley.com/10.1002/qj.2401}.

\bibitem[{Hansen and Penland(2007)Hansen, and Penland}]{Hansen2007a}
Hansen, J.~A., and C.~Penland, 2007: {On stochastic parameter estimation using
  data assimilation}. \textit{Physica D: Nonlinear Phenomena},
  \textbf{230~(1-2)}, 88--98, \doi{10.1016/j.physd.2006.11.006}.

\bibitem[{Hasselmann(1988)}]{Hasselman1988}
Hasselmann, K., 1988: {PIPs and POPs: The reduction of complex dynamical
  systems using principal interaction and oscillation patterns}.
  \textit{Journal of Geophysical Research}, \textbf{93~(D9)}, 11\,015,
  \doi{10.1029/JD093iD09p11015},
  \urlprefix\url{http://doi.wiley.com/10.1029/JD093iD09p11015}.

\bibitem[{Hastie et~al.(2017)Hastie, Tibshirani,, and Friedman}]{Hastie2017}
Hastie, T., R.~Tibshirani, and J.~Friedman, 2017: \textit{{The Elements of
  Statistical Learning Data Mining, Inference, and Prediction (12th
  printing)}}. 2nd ed., Springer-Verlag New York, New York, 745 pp.,
  \doi{10.1007/978-0-387-84858-7},
  \urlprefix\url{http://link.springer.com/10.1007/978-0-387-84858-7}.

\bibitem[{Hawkins et~al.(2011)Hawkins, Robson, Sutton, Smith,, and
  Keenlyside}]{hawkins2011evaluating}
Hawkins, E., J.~Robson, R.~Sutton, D.~Smith, and N.~Keenlyside, 2011:
  {Evaluating the potential for statistical decadal predictions of sea surface
  temperatures with a perfect model approach}. \textit{Climate Dynamics},
  \textbf{37~(11-12)}, 2495--2509, \doi{10.1007/s00382-011-1023-3},
  \urlprefix\url{http://link.springer.com/10.1007/s00382-011-1023-3}.

\bibitem[{Hawkins and Sutton(2009)Hawkins, and Sutton}]{Hawkins2009}
Hawkins, E., and R.~Sutton, 2009: {Decadal predictability of the Atlantic Ocean
  in a coupled GCM: Forecast skill and optimal perturbations using linear
  inverse modeling}. \textit{Journal of Climate}, \textbf{22~(14)}, 3960--3978,
  \doi{10.1175/2009JCLI2720.1},
  \urlprefix\url{http://journals.ametsoc.org/doi/abs/10.1175/2009JCLI2720.1}.

\bibitem[{Jenkins et~al.(2016)Jenkins, Dutrieux, Jacobs, Steig, Gudmundsson,
  Smith,, and Heywood}]{jenkins2016decadal}
Jenkins, A., P.~Dutrieux, S.~Jacobs, E.~J. Steig, G.~H. Gudmundsson, J.~Smith,
  and K.~J. Heywood, 2016: {Decadal ocean forcing and Antarctic ice sheet
  response: Lessons from the Amundsen Sea}. \textit{Oceanography},
  \textbf{29~(4)}, 106--117, \doi{10.5670/oceanog.2016.103},
  \urlprefix\url{https://tos.org/oceanography/article/decadal-ocean-forcing-and-antarctic-ice-sheet-response-lessons-from-the-amu}.

\bibitem[{Karimi and McAuley(2016)Karimi, and McAuley}]{Karimi2016}
Karimi, H., and K.~B. McAuley, 2016: {Bayesian Estimation in Stochastic
  Differential Equation Models via Laplace Approximation}.
  \textit{IFAC-PapersOnLine}, \textbf{49~(7)}, 1109--1114,
  \doi{10.1016/j.ifacol.2016.07.351},
  \urlprefix\url{http://folk.ntnu.no/skoge/prost/proceedings/dycops-cab-2016/proceedings/media/papers/0007.pdf}.

\bibitem[{Kay et~al.(2015)}]{EKaya}
Kay, J.~E., and Coauthors, 2015: {The community earth system model (CESM) large
  ensemble project : A community resource for studying climate change in the
  presence of internal climate variability}. \textit{Bulletin of the American
  Meteorological Society}, \textbf{96~(8)}, 1333--1349,
  \doi{10.1175/BAMS-D-13-00255.1}, \urlprefix\url{http://journals.ametsoc}.

\bibitem[{Kondrashov et~al.(2015)Kondrashov, Chekroun,, and
  Ghil}]{Kondrashov2015a}
Kondrashov, D., M.~D. Chekroun, and M.~Ghil, 2015: {Data-driven non-Markovian
  closure models}. \textit{Physica D: Nonlinear Phenomena}, \textbf{297},
  33--55, \doi{10.1016/j.physd.2014.12.005}.

\bibitem[{Kravtsov et~al.(2009)Kravtsov, Kondrashov,, and Ghil}]{Kravtsov2009a}
Kravtsov, S., D.~Kondrashov, and M.~Ghil, 2009: {Empirical model reduction and
  the modelling hierarchy in climate dynamics and the geosciences}.
  \textit{Stochastic Physics and Climate Modelling}, 35--72,
  \urlprefix\url{http://www.atmos.ucla.edu/tcd/PREPRINTS/BookEMR_Text.pdf}.

\bibitem[{Lachlan-Cope and Connolley(2006)Lachlan-Cope, and
  Connolley}]{lachlan2006teleconnections}
Lachlan-Cope, T., and W.~Connolley, 2006: {Teleconnections between the tropical
  Pacific and the Amundsen-Bellinghausens Sea: Role of the El
  Ni{\~{n}}o/Southern Oscillation}. \textit{Journal of Geophysical Research
  Atmospheres}, \textbf{111~(23)}, n/a--n/a, \doi{10.1029/2005JD006386},
  \urlprefix\url{http://doi.wiley.com/10.1029/2005JD006386}.

\bibitem[{Latif et~al.(2017)Latif, Martin, Reintges,, and
  Park}]{latif2017southern}
Latif, M., T.~Martin, A.~Reintges, and W.~Park, 2017: {Southern Ocean Decadal
  Variability and Predictability}. Springer International Publishing,
  \urlprefix\url{http://link.springer.com/10.1007/s40641-017-0068-8}, 163--173
  pp., \doi{10.1007/s40641-017-0068-8}.

\bibitem[{Le~Breton(1977)}]{LeBreton1977}
Le~Breton, A., 1977: {Parameter Estimation in a Linear Stochastic Differential
  Equation}. \textit{Transactions of the Seventh Prague Conference on
  Information Theory, Statistical Decision Functions, Random Processes and of
  the 1974 European Meeting of Statisticians}, Springer Netherlands, Dordrecht,
  353--366, \doi{10.1007/978-94-010-9910-3{\_}36},
  \urlprefix\url{http://link.springer.com/10.1007/978-94-010-9910-3_36}.

\bibitem[{Lewandowski et~al.(2009)Lewandowski, Kurowicka,, and
  Joe}]{Lewandowski2009}
Lewandowski, D., D.~Kurowicka, and H.~Joe, 2009: {Generating random correlation
  matrices based on vines and extended onion method}. \textit{Journal of
  Multivariate Analysis}, \textbf{100~(9)}, 1989--2001,
  \doi{10.1016/j.jmva.2009.04.008},
  \urlprefix\url{https://www.sciencedirect.com/science/article/pii/S0047259X09000876}.

\bibitem[{Litterman(1986)}]{Litterman1986a}
Litterman, R.~B., 1986: {Forecasting with bayesian vector
  autoregressions—five years of experience}. \textit{Journal of Business and
  Economic Statistics}, \textbf{4~(1)}, 25--38,
  \doi{10.1080/07350015.1986.10509491},
  \urlprefix\url{https://digidownload.libero.it/rocco.mosconi/Litterman1986.pdf}.

\bibitem[{Majda and Harlim(2013)Majda, and Harlim}]{Majda}
Majda, A.~J., and J.~Harlim, 2013: {Physics constrained nonlinear regression
  models for time series}. \textit{Nonlinearity}, \textbf{26~(1)}, 201--217,
  \doi{10.1088/0951-7715/26/1/201},
  \urlprefix\url{https://www.math.nyu.edu/faculty/majda/Submitted/Majda Harlim
  (2012) Physics constrained.pdf}.

\bibitem[{Martinez-Villalobos et~al.(2019)Martinez-Villalobos, Newman, Vimont,
  Penland,, and David~Neelin}]{MartinezVillalobos2019}
Martinez-Villalobos, C., M.~Newman, D.~J. Vimont, C.~Penland, and
  J.~David~Neelin, 2019: {Observed El Ni{\~{n}}o-La Ni{\~{n}}a Asymmetry in a
  Linear Model}. \textit{Geophysical Research Letters}, \textbf{46~(16)},
  9909--9919, \doi{10.1029/2019GL082922},
  \urlprefix\url{https://onlinelibrary.wiley.com/doi/abs/10.1029/2019GL082922}.

\bibitem[{Martinez-Villalobos et~al.(2018)Martinez-Villalobos, Vimont, Penland,
  Newman,, and Neelin}]{Martinez2018}
Martinez-Villalobos, C., D.~J. Vimont, C.~Penland, M.~Newman, and J.~D. Neelin,
  2018: {Calculating state-dependent noise in a linear inverse model
  framework}. \textit{Journal of the Atmospheric Sciences}, \textbf{75~(2)},
  479--496, \doi{10.1175/JAS-D-17-0235.1},
  \urlprefix\url{https://doi.org/10.1175/JAS-D-17-
  http://journals.ametsoc.org/doi/10.1175/JAS-D-17-0235.1}.

\bibitem[{Mason et~al.(2002)Mason, Mimmack, Mason,, and Mimmack}]{Mason2002}
Mason, S.~J., G.~M. Mimmack, S.~J. Mason, and G.~M. Mimmack, 2002: {Comparison
  of Some Statistical Methods of Probabilistic Forecasting of ENSO}.
  \textit{Journal of Climate}, \textbf{15~(1)}, 8--29,
  \doi{10.1175/1520-0442(2002)015<0008:COSSMO>2.0.CO;2},
  \urlprefix\url{http://journals.ametsoc.org/doi/abs/10.1175/1520-0442%282002%29015%3C0008%3ACOSSMO%3E2.0.CO%3B2}.

\bibitem[{McDermott and Wikle(2019{\natexlab{a}})McDermott, and
  Wikle}]{McDermott2019}
McDermott, P.~L., and C.~K. Wikle, 2019{\natexlab{a}}: {Bayesian recurrent
  neural network models for forecasting and quantifying uncertainty in
  spatial-temporal data}. \textit{Entropy}, \textbf{21~(2)},
  \doi{10.3390/e21020184}, \urlprefix\url{http://arxiv.org/abs/1711.00636}.

\bibitem[{McDermott and Wikle(2019{\natexlab{b}})McDermott, and
  Wikle}]{McDermott2019a}
McDermott, P.~L., and C.~K. Wikle, 2019{\natexlab{b}}: {Deep echo state
  networks with uncertainty quantification for spatio-temporal forecasting}.
  \textit{Environmetrics}, \textbf{30~(3)}, \doi{10.1002/env.2553},
  \urlprefix\url{http://arxiv.org/abs/1806.10728}.

\bibitem[{McKinnon and Deser(2018)McKinnon, and Deser}]{McKinnon2018}
McKinnon, K.~A., and C.~Deser, 2018: {Internal variability and regional climate
  trends in an observational large ensemble}. \textit{Journal of Climate},
  \textbf{31~(17)}, 6783--6802, \doi{10.1175/JCLI-D-17-0901.1},
  \urlprefix\url{http://journals.ametsoc.org/doi/10.1175/JCLI-D-17-0901.1}.

\bibitem[{Meehl et~al.(2014)}]{meehl2014decadal}
Meehl, G.~A., and Coauthors, 2014: {Decadal climate prediction an update from
  the trenches}. \textit{Bulletin of the American Meteorological Society},
  \textbf{95~(2)}, 243--267, \doi{10.1175/BAMS-D-12-00241.1},
  \urlprefix\url{http://journals.ametsoc.org/doi/abs/10.1175/BAMS-D-12-00241.1}.

\bibitem[{M{\o}ller et~al.(2011)M{\o}ller, Madsen,, and Carstensen}]{Mller2011}
M{\o}ller, J.~K., H.~Madsen, and J.~Carstensen, 2011: {Parameter estimation in
  a simple stochastic differential equation for phytoplankton modelling}.
  \textit{Ecological Modelling}, \textbf{222}, 1793--1799,
  \doi{10.1016/j.ecolmodel.2011.03.025},
  \urlprefix\url{http://henrikmadsen.org/wp-content/uploads/2014/05/Journal_article_-_2011_-_Parameter_estimation_in_a_simple_stochastic_differential_equation_for_phytoplankton_modelling.pdf}.

\bibitem[{Neal(2012)}]{Neal2012}
Neal, R.~M., 2012: {MCMC using Hamiltonian dynamics}. \doi{10.1201/b10905-6},
  \urlprefix\url{https://arxiv.org/pdf/1206.1901.pdf
  http://arxiv.org/abs/1206.1901}.

\bibitem[{Newman(2007)}]{Newman2007}
Newman, M., 2007: {Interannual to decadal predictability of tropical and North
  Pacific sea surface temperatures}. \textit{Journal of Climate},
  \textbf{20~(11)}, 2333--2356, \doi{10.1175/JCLI4165.1},
  \urlprefix\url{http://journals.ametsoc.org/doi/abs/10.1175/JCLI4165.1}.

\bibitem[{Newman(2013)}]{Newman2013a}
Newman, M., 2013: {An Empirical Benchmark for Decadal Forecasts of Global
  Surface Temperature Anomalies}. \textit{Journal of Climate},
  \textbf{26~(14)}, 5260--5269, \doi{10.1175/JCLI-D-12-00590.1},
  \urlprefix\url{http://journals.ametsoc.org/doi/abs/10.1175/JCLI-D-12-00590.1}.

\bibitem[{Nummelin et~al.(2018)Nummelin, Jeffress,, and Haine}]{Nummelin2018}
Nummelin, A., S.~Jeffress, and T.~Haine, 2018: {Statistical inversion of
  surface ocean kinematics from sea surface temperature observations}.
  \textit{Journal of Atmospheric and Oceanic Technology}, \textbf{35~(10)},
  1913--1933, \doi{10.1175/JTECH-D-18-0057.1},
  \urlprefix\url{http://journals.ametsoc.org/doi/10.1175/JTECH-D-18-0057.1}.

\bibitem[{Partridge and Rickman(1998)Partridge, and Rickman}]{Partridge1998}
Partridge, M.~D., and D.~S. Rickman, 1998: {Generalizing the Bayesian Vector
  Autoregression Approach for Regional Interindustry Employment Forecasting}.
  \textit{Journal of Business {\&} Economic Statistics}, \textbf{16~(1)}, 62,
  \doi{10.2307/1392016},
  \urlprefix\url{https://www.jstor.org/stable/1392016?origin=crossref}.

\bibitem[{Penland(1989)}]{Penland1989}
Penland, C., 1989: {Random forcing and forecasting using principal oscillation
  pattern analysis}. \textit{Monthly Weather Review}, \textbf{117~(10)},
  2165--2185, \doi{10.1175/1520-0493(1989)117<2165:RFAFUP>2.0.CO;2},
  \urlprefix\url{http://journals.ametsoc.org/doi/abs/10.1175/1520-0493%281989%29117%3C2165%3ARFAFUP%3E2.0.CO%3B2}.

\bibitem[{Penland(1996)}]{Penland1996}
Penland, C., 1996: {A stochastic model of IndoPacific sea surface temperature
  anomalies}. Tech. Rep. 2-4, 534--558 pp. \doi{10.1016/0167-2789(96)00124-8},
  \urlprefix\url{https://ac.els-cdn.com/0167278996001248/1-s2.0-0167278996001248-main.pdf?_tid=5bab1b2a-8e9d-4920-aeb9-45e5ab3eb1cb&acdnat=1541143825_3408dccbfcca2527b42894a255fa14f0}.

\bibitem[{Penland et~al.(1993)Penland, Magorian, Penland,, and
  Magorian}]{Penland1993}
Penland, C., T.~Magorian, C.~Penland, and T.~Magorian, 1993: {Prediction of
  Ni{\~{n}}o 3 Sea Surface Temperatures Using Linear Inverse Modeling}.
  \textit{Journal of Climate}, \textbf{6~(6)}, 1067--1076,
  \doi{10.1175/1520-0442(1993)006<1067:PONSST>2.0.CO;2},
  \urlprefix\url{http://journals.ametsoc.org/doi/abs/10.1175/1520-0442%281993%29006%3C1067%3APONSST%3E2.0.CO%3B2}.

\bibitem[{Penland and Matrosova(2001)Penland, and Matrosova}]{Penland2001}
Penland, C., and L.~Matrosova, 2001: {Expected and Actual Errors of Linear
  Inverse Model Forecasts}. \textit{Monthly Weather Review},
  \doi{10.1016/0306-4522(81)90148-2}.

\bibitem[{Penland and Sardeshmukh(1995)Penland, and Sardeshmukh}]{Penland1995}
Penland, C., and P.~D. Sardeshmukh, 1995: {The Optimal Growth of Tropical Sea
  Surface Temperature Anomalies}. \textit{Journal of Climate}, \textbf{8~(8)},
  1999--2024, \doi{10.1175/1520-0442(1995)008<1999:TOGOTS>2.0.CO;2},
  \urlprefix\url{http://journals.ametsoc.org/doi/abs/10.1175/1520-0442%281995%29008%3C1999%3ATOGOTS%3E2.0.CO%3B2}.

\bibitem[{R{\'{e}}velard et~al.(2018)R{\'{e}}velard, Frankignoul,, and
  Kwon}]{Revelard2018}
R{\'{e}}velard, A., C.~Frankignoul, and Y.~O. Kwon, 2018: {A multivariate
  estimate of the cold season atmospheric response to North Pacific SST
  variability}. \textit{Journal of Climate}, \textbf{31~(7)}, 2771--2796,
  \doi{10.1175/JCLI-D-17-0061.1},
  \urlprefix\url{http://journals.ametsoc.org/doi/10.1175/JCLI-D-17-0061.1}.

\bibitem[{Samaniego(2010)}]{Samaniego2010}
Samaniego, F.~J., 2010: \textit{{A Comparison of the Bayesian and Frequentist
  Approaches to Estimation}}. 1st ed., Springer-Verlag New York, New York,
  XIII, 225 pp., \doi{10.1007/978-1-4419-5941-6}.

\bibitem[{Sardeshmukh et~al.(2015)Sardeshmukh, Compo,, and
  Penland}]{Sardeshmukh2015}
Sardeshmukh, P.~D., G.~P. Compo, and C.~Penland, 2015: {Need for caution in
  interpreting extreme weather statistics}. \textit{Journal of Climate},
  \textbf{28~(23)}, 9166--9187, \doi{10.1175/JCLI-D-15-0020.1},
  \urlprefix\url{http://journals.ametsoc.org/doi/10.1175/JCLI-D-15-0020.1}.

\bibitem[{S{\"{a}}rkk{\"{a}}(2010)}]{Sarkka2013}
S{\"{a}}rkk{\"{a}}, S., 2010: \textit{{Bayesian filtering and smoothing}}.
  Cambridge University Press, Cambridge, 1--232 pp.,
  \doi{10.1017/CBO9781139344203},
  \urlprefix\url{http://ebooks.cambridge.org/ref/id/CBO9781139344203}.

\bibitem[{Steig et~al.(2012)Steig, Ding, Battisti,, and
  Jenkins}]{steig2012tropical}
Steig, E.~J., Q.~Ding, D.~S. Battisti, and A.~Jenkins, 2012: {Tropical forcing
  of circumpolar deep water inflow and outlet glacier thinning in the amundsen
  sea embayment, west antarctica}. \textit{Annals of Glaciology},
  \textbf{53~(60)}, 19--28, \doi{10.3189/2012AoG60A110},
  \urlprefix\url{https://www.cambridge.org/core/product/identifier/S0260305500251690/type/journal_article}.

\bibitem[{Stock et~al.(2015{\natexlab{a}})}]{Singer}
Stock, C.~A., and Coauthors, 2015{\natexlab{a}}: {a Survey of Estimation
  Methods for Stochastic Differential Equations}. \textit{Differential
  Equations}, \textbf{137~(3)}, 1--12, \doi{10.1016/j.pocean.2015.06.007},
  \urlprefix\url{https://core.ac.uk/download/pdf/11559876.pdf}.

\bibitem[{Stock et~al.(2015{\natexlab{b}})}]{Stock2015}
Stock, C.~A., and Coauthors, 2015{\natexlab{b}}: {Seasonal sea surface
  temperature anomaly prediction for coastal ecosystems}. \textit{Progress in
  Oceanography}, \textbf{137}, 219--236, \doi{10.1016/j.pocean.2015.06.007},
  \urlprefix\url{http://dx.doi.org/10.1016/j.pocean.2015.06.007}.

\bibitem[{Stoer and Bulirsch(2002)Stoer, and Bulirsch}]{Stoer2002a}
Stoer, J., and R.~Bulirsch, 2002: {Eigenvalue Problems}. \textit{Introduction
  to Numerical Analysis}, Springer New York, New York, NY, 364--464,
  \doi{10.1007/978-0-387-21738-3{\_}6},
  \urlprefix\url{http://link.springer.com/10.1007/978-0-387-21738-3_6}.

\bibitem[{Thomas et~al.(2018)Thomas, Vimont, Newman, Penland,, and
  Mart{\'{i}}nez-Villalobos}]{Thomas2018}
Thomas, E.~E., D.~J. Vimont, M.~Newman, C.~Penland, and
  C.~Mart{\'{i}}nez-Villalobos, 2018: {The role of stochastic forcing in
  generating ENSO diversity}. \textit{Journal of Climate}, \textbf{31~(22)},
  9125--9150, \doi{10.1175/JCLI-D-17-0582.1},
  \urlprefix\url{http://journals.ametsoc.org/doi/10.1175/JCLI-D-17-0582.1}.

\bibitem[{Tian et~al.(2014)Tian, Zhou, Wu,, and Ge}]{Tian2014}
Tian, T., Y.~Zhou, Y.~Wu, and X.~Ge, 2014: {Estimation of Parameters in
  Mean-Reverting Stochastic Systems}. \textit{Mathematical Problems in
  Engineering}, \textbf{2014}, 1--8, \doi{10.1155/2014/317059},
  \urlprefix\url{http://www.hindawi.com/journals/mpe/2014/317059/}.

\bibitem[{van Zanten(2013)}]{VanZanten2013}
van Zanten, H., 2013: {Nonparametric Bayesian methods for one-dimensional
  diffusion models}. \textit{Mathematical Biosciences}, \textbf{243~(2)},
  215--222, \doi{10.1016/j.mbs.2013.03.008},
  \urlprefix\url{https://arxiv.org/pdf/1209.6433.pdf}.

\bibitem[{Vogel(2002)}]{Vogel2002}
Vogel, C.~R., 2002: \textit{{Computational Methods for Inverse Problems}}.
  Society for Industrial and Applied Mathematics,
  \doi{10.1137/1.9780898717570},
  \urlprefix\url{https://epubs.siam.org/doi/abs/10.1137/1.9780898717570}.

\bibitem[{von Storch and Bruns(1988)von Storch, and Bruns}]{vonStorch1988}
von Storch, H., and T.~Bruns, 1988: {Principal oscillation pattern analysis of
  the 30‐to 60‐day oscillation in general circulation model equatorial
  troposphere}. \textit{Journal of Geo}, \textbf{93~(D9)}, 11\,022--11\,036,
  \doi{10.1029/JD093iD09p11022},
  \urlprefix\url{http://doi.wiley.com/10.1029/JD093iD09p11022
  http://onlinelibrary.wiley.com/doi/10.1029/JD093iD09p11022/full}.

\bibitem[{Wikle and Holan(2011)Wikle, and Holan}]{Wikle2011}
Wikle, C.~K., and S.~H. Holan, 2011: {Polynomial nonlinear spatio-temporal
  integro-difference equation models}. \textit{Journal of Time Series
  Analysis}, \textbf{32~(4)}, 339--350, \doi{10.1111/j.1467-9892.2011.00729.x},
  \urlprefix\url{http://doi.wiley.com/10.1111/j.1467-9892.2011.00729.x}.

\bibitem[{Wittenberg et~al.(2014)Wittenberg, Rosati, Delworth, Vecchi,, and
  Zeng}]{Wittenberg2014}
Wittenberg, A.~T., A.~Rosati, T.~L. Delworth, G.~A. Vecchi, and F.~Zeng, 2014:
  {ENSO modulation: Is it decadally predictable?} \textit{Journal of Climate},
  \textbf{27~(7)}, 2667--2681, \doi{10.1175/JCLI-D-13-00577.1},
  \urlprefix\url{http://journals.ametsoc.org/doi/abs/10.1175/JCLI-D-13-00577.1}.

\bibitem[{Wu et~al.(2018)Wu, Notaro, Vavrus, Mortensen, Montgomery,
  de~Pi{\'{e}}rola,, and Block}]{Wu2018}
Wu, S., M.~Notaro, S.~Vavrus, E.~Mortensen, R.~Montgomery, J.~de~Pi{\'{e}}rola,
  and P.~Block, 2018: {Efficacy of tendency and linear inverse models to
  predict southern Peru's rainy season precipitation}. \textit{International
  Journal of Climatology}, \textbf{38~(5)}, 2590--2604, \doi{10.1002/joc.5442},
  \urlprefix\url{http://doi.wiley.com/10.1002/joc.5442}.

\bibitem[{Zhang et~al.(2017)Zhang, Delworth,, and Jia}]{zhang2017diagnosis}
Zhang, L., T.~L. Delworth, and L.~Jia, 2017: {Diagnosis of decadal
  predictability of Southern Ocean sea surface temperature in the GFDL CM2.1
  Model}. \textit{Journal of Climate}, \textbf{30~(16)}, 6309--6328,
  \doi{10.1175/JCLI-D-16-0537.1},
  \urlprefix\url{http://journals.ametsoc.org/doi/10.1175/JCLI-D-16-0537.1}.

\bibitem[{Zhao et~al.(2019)Zhao, Liu, Zheng,, and Jin}]{Zhao2019}
Zhao, Y., Z.~Liu, F.~Zheng, and Y.~Jin, 2019: {Parameter optimization for
  real-world ENSO forecast in an intermediate coupled model}. \textit{Monthly
  Weather Review}, \textbf{147~(5)}, 1429--1445, \doi{10.1175/MWR-D-18-0199.1},
  \urlprefix\url{www.ametsoc.org/PUBSReuseLicenses}.

\end{thebibliography}



\end{document}